\documentclass[prd,superscriptaddress,altaffilletter,nofootinbib]{revtex4}

\usepackage{cancel}
\usepackage{graphicx}
\usepackage{amsmath}
\usepackage{amssymb}
\usepackage{graphicx,epsfig}
\usepackage[dvipsnames]{xcolor}
\newcommand{\be}{\begin{equation}}
\newcommand{\ee}{\end{equation}}
\newcommand{\bea}{\begin{eqnarray}}
\newcommand{\eea}{\end{eqnarray}}
\newcommand{\der}{\partial}
\newcommand{\vphi}{\varphi}

\begin{document}



\title{Could a so far ignored symmetry of the classical laws of gravity explain the cosmological puzzles?}
 

\author{Israel Quiros}\email{i.quiros@ugto.mx; iquiros@fisica.ugto.mx}\affiliation{Departamento Ingenier\'ia Civil, Divisi\'on de Ingenier\'ia, Universidad de Guanajuato, C.P. 36000, Gto., M\'exico.}



\date{\today}


\begin{abstract} We show that if the masses of timelike fields are point-dependent quantities transforming under conformal transformations as $m\rightarrow\Omega^{-1}m$, so the energy density of perfect fluids transforms as $\rho\rightarrow\Omega^{-4}\rho$, form-invariance under Weyl transformations could be an actual symmetry of the gravitational interactions of matter. That is, under the mentioned circumstances, Weyl symmetry allows any matter field to be coupled to gravity. The phenomenological and physical consequences of the novel result, including the ``many worlds'' interpretation of gauge freedom, are drawn. We explore, in particular, a possible explanation of two major cosmological puzzles: dark matter and dark energy, as a consequence of Weyl symmetry. Quantum-mechanical removal of the spacetime singularities in this framework is briefly discussed as well.\end{abstract}





\maketitle




\section{Introduction}
\label{sect-intro}   


Conformal symmetry is perhaps the least understood symmetry in physics \cite{dicke-1962, deser-1970, callan_prd_1970,  morganstern_1970, rosen-1970, anderson-1971, fujii-1974, bekenstein_1980, cheng-1988, bekenstein_1993, magnano_1994, capozziello_1997, faraoni_1998, faraoni_rev, faraoni_ijmpd_1999, quiros_prd_2000, fabris_2000, fujii_book, faraoni_book, flanagan_cqg_2004, faraoni_prd_2007, catena_prd_2007, nicolai-2007, sotiriou_ijmpd_2008, odi-1, elizalde_grg_2010, deruelle_2011, chiba_2013, capozziello_prd_2013, quiros_grg_2013, bars-2014, alvarez-2015, jarv_2015, jackiw-2015, sasaki_2016, banerjee_2016, pandey_2017, quiros_ijmpd_2019, hobson-prd-2020, hobson-epjc-2022, bamber_prd_2023, mohamedi-2024, quiros_prd_2025}. On the one hand, it has been claimed in \cite{dicke-1962}, and since then widely accepted as a physical principle \cite{faraoni_prd_2007}, that the laws of physics must be invariant under conformal transformations (CTs), also known as ``Weyl rescalings,'' ``local scale transformations'' (LSTs), or ``transformations of units'' \cite{bekenstein_1980, faraoni_rev, fujii_book, faraoni_book, faraoni_prd_2007, quiros_ijmpd_2019, hobson-prd-2020, hobson-epjc-2022, casas_1992};

\begin{align} g_{\mu\nu}\rightarrow\hat g_{\mu\nu}=\Omega^2g_{\mu\nu},\;\Phi_i\rightarrow\hat\Phi_i=\Omega^{w_i}\Phi_i,\label{conf-t}\end{align} where $\Omega=\Omega(x)$ is the conformal factor ($\Omega(x)>0$), $\Phi_i$ are other gravitational and matter fields present, and $w_i$ are the conformal weights of these fields. Here we call this invariance ``Dicke's principle.''

The high-energy physics (HEP) argument that, due to the Planck mass $M_\text{pl}$, being a dimensionful parameter, gravity cannot be conformally invariant, seems to oppose Dicke's principle \cite{dicke-1962}. In general, the presence of any mass term is believed to break the conformal invariance \cite{deser-1970, callan_prd_1970}. However, strictly speaking, this is true only if we consider that constant masses do not transform under \eqref{conf-t}: $m\rightarrow m$. As stated in \cite{bekenstein_1980}, we must differentiate between LSTs:

\begin{align} g_{\mu\nu}\rightarrow\Omega^2g_{\mu\nu},\;\Phi_i\rightarrow\Omega^{w_i}\Phi_i,\;m\rightarrow m,\label{scale-t}\end{align} and Weyl transformations (WTs) (same as transformations of units \cite{dicke-1962}):

\begin{align} g_{\mu\nu}\rightarrow\Omega^2g_{\mu\nu},\;\Phi_i\rightarrow\Omega^{w_i}\Phi_i,\;m\rightarrow\Omega^{-1}m,\label{units-t}\end{align} where, in addition to CTs \eqref{conf-t}, the transformation of the masses $m\rightarrow\Omega^{-1}m$, is assumed. That is, the mass parameters are fields with conformal weight $w_m=-1$ \cite{hobson-prd-2020, hobson-epjc-2022, casas_1992}. According to \cite{bekenstein_1980}, scale transformations \eqref{scale-t} are not unit transformations; they are active enlargements of a system whose usefulness depends on the absence of a length scale. Meanwhile, WTs \eqref{units-t} are units' transformations whose physical and geometrical meanings follow from the existence of some scale of length.


The transformation of the mass $m$ and of the Planck constant $h$, under Weyl transformations, are not independent of each other. The Planck constant has the following dimensions: $[h]=[M][L]^2[T]^{-1}$, where $[M]$ is the mass dimension, while $[L]$ and $[T]$ are the length and time dimensions, respectively. Under WTs, the latter dimensions transform as the proper length element $dl=\sqrt{g_{ik}dx^idx^k}$ and the proper time element $d\tau=\sqrt{g_{00}dt^2}$ do. Consequently, $[L]^2[T]^{-1}\rightarrow\Omega\,[L]^2[T]^{-1}$. Hence, if we assume that the CTs do not transform the Planck constant, the mass parameter $m$ must be transformed according to \eqref{units-t}. Otherwise, if we assume that the mass parameter is a constant, as in \cite{deser-1970}, then the Planck constant transforms as $h\rightarrow\Omega\,h$. Regardless of whether the mass or the Planck constant is transformed under CTs, the Compton wavelength $\lambda_c=h/m$ transforms as any length scale $\lambda_c\to\Omega\,\lambda_c$.


In the present document, as is customary within the framework of scalar-tensor theories (STTs) \cite{dicke-1962, bekenstein_1980, faraoni_rev, fujii_book, faraoni_book, faraoni_prd_2007, quiros_ijmpd_2019, hobson-prd-2020, hobson-epjc-2022, casas_1992}, we consider WTs \eqref{units-t}, instead of LSTs \eqref{scale-t}. In consequence, it is assumed that the masses are point-dependent fields that, under \eqref{conf-t}, transform like in \eqref{units-t}. 

Whether the mass parameter transforms or not under CTs plays a central role in the demonstration of conformal form-invariance of the action of fundamental matter fields \cite{quiros_prd_2025}:

\begin{align} S_m=\int d^4x\,{\cal L}_m(\chi_i,\der\chi_i,g_{\mu\nu})=\int d^4x\sqrt{-g}\,L_m(\chi_i,\der\chi_i,g_{\mu\nu}),\label{mat-action}\end{align} where $L_m(\chi_i,\der\chi_i,g_{\mu\nu})$ is the Lagrangian of the matter fields $\chi_i$, with $i=1,2,3,...,N$, and $N$ -- the number of matter fields that are minimally coupled to the metric $g_{\mu\nu}$, while ${\cal L}_m(\chi_i,\der\chi_i,g_{\mu\nu})$ is the corresponding Lagrangian density. Here, the stress-energy tensor (SET) of the matter fields is defined as usual;

\begin{align} T^{(m)}_{\mu\nu}=-\frac{2}{\sqrt{-g}}\frac{\delta{\cal L}_m}{\delta g^{\mu\nu}}.\label{mat-set}\end{align} 


For fermions and gauge bosons, it has been shown in \cite{quiros_prd_2025} that if their masses are point-dependent quantities transforming as in \eqref{units-t}, the matter action \eqref{mat-action} is form-invariant under WTs

\begin{align} g_{\mu\nu}\rightarrow\hat g_{\mu\nu}=\Omega^2g_{\mu\nu},\;\chi_i\rightarrow\hat\chi_i=\Omega^{w_{\chi_i}}\chi_i,\;m\rightarrow\hat m=\Omega^{-1}m.\label{gauge-t'}\end{align} But if the masses are constant parameters, then \eqref{mat-action} is not form-invariant under \eqref{scale-t}. Take, for illustration, the Lagrangian density of the Dirac fermion:\footnote{We use the following signature of the metric: $(-,+,+,+)$, and follow the notation of \cite{quiros_prd_2025}.} 

\begin{align} {\cal L}_\text{fermion}=\sqrt{-g}\,\bar\psi\left(i\cancel{\cal D}-m_\psi\right)\psi,\label{fermion-lag}\end{align} where $(\psi,\bar\psi)$ are the fermion spinor and its adjoint spinor. The massless piece $\sqrt{-g}\bar\psi\cancel{\cal D}\psi$ is form-invariant under simultaneous Weyls transformations \eqref{conf-t} and $\psi\rightarrow\Omega^{-3/2}\psi$ ($\bar\psi\rightarrow\Omega^{-3/2}\bar\psi$) since $\cancel{\cal D}\psi\rightarrow\Omega^{-5/2}\cancel{\cal D}\psi$ and $\sqrt{-g}\rightarrow\Omega^4\sqrt{-g}$. However, form-invariance of the matter term $\sqrt{-g}\,m_\psi\bar\psi\psi$, depends on the transformation property of the mass parameter $m_\psi$ under CTs. If the mass is an untransformed constant, then $\sqrt{-g}\,m_\psi\bar\psi\psi\rightarrow\Omega\sqrt{-g}\,m_\psi\bar\psi\psi$, so the mass piece of Dirac's Lagrangian density breaks conformal invariance. In contrast, if the mass $m$ is a point-dependent quantity that transforms as in \eqref{units-t} under CTs, then the matter action \eqref{mat-action} is conformal form-invariant for any fundamental (timelike) matter fields. In the above example, if the mass $m_\psi$ transforms as $m_\psi\rightarrow\Omega^{-1}m_\psi$, then the mass piece of the Lagrangian density of the Dirac fermion is form-invariant under CTs: $\sqrt{-g}\,m_\psi\bar\psi\psi\rightarrow\sqrt{-g}\,m_\psi\bar\psi\psi$ and so is the total Lagrangian density \eqref{fermion-lag}. Hence, form-invariance of the matter action under \eqref{gauge-t'} depends on which transformations are considered: Weyl transformations or local scale transformations.


A large body of applications of gravitational theories is associated with background matter in the form of a perfect fluid, rather than with fundamental fields. Therefore, it is of interest to investigate the transformation properties of the action of perfect fluids under Weyl transformations \eqref{units-t}. In \cite{quiros_prd_2025}, it has also been shown that, in addition to the action of fundamental matter fields, that of perfect fluids is form-invariant under WTs as well. Actually, the action of a perfect fluid with energy density $\rho$ and barotropic pressure $p$ can be written in either form: $S_\text{fluid}=-\int d^4x\sqrt{-g}\,\rho$ or $S_\text{fluid}=\int d^4x\sqrt{-g}\,p$. Unless the perfect fluid couples explicitly to the curvature (which is not the case in this paper), both Lagrangian densities are equivalent \cite{faraoni_2009}. Here, for definiteness, we choose the former action for the perfect fluid. According to dimensional arguments, if we assume point-dependent masses that transform as $m\rightarrow\Omega^{-1}m$ under CTs \eqref{conf-t}, the energy density of the fluid transforms as $\rho\rightarrow\Omega^{-4}\rho$. In fact, the energy density has units: $[M]/[L]^3$, where $[M]\rightarrow\Omega^{-1}[M]$ and $[L]\rightarrow\Omega[L]$ are the transformation laws of the units of mass and length under CTs, respectively. Hence, $[M]/[L]^3\rightarrow\Omega^{-4}[M]/[L]^3$. The barotropic pressure of the fluid $p$ transforms in the same way: $p\rightarrow\Omega^{-4}p$. Since under \eqref{conf-t}, $\sqrt{-g}\rightarrow\sqrt{-\hat g}=\Omega^4\sqrt{-g}$ and $\rho\rightarrow\hat\rho=\Omega^{-4}\rho$, then $\int d^4x\sqrt{-g}\,\rho=\int d^4x\sqrt{-\hat g}\,\hat\rho$. That is, the action of the perfect fluid is form-invariant under WTs. Note, however, that if we assume that the mass unit is not transformed under CTs, then $\rho\rightarrow\hat\rho=\Omega^{-3}\rho$, which means that the action of the perfect fluid is not form-invariant: $\int d^4x\sqrt{-g}\,\rho=\int d^4x\,\Omega\sqrt{-\hat g}\hat\rho$.


The requirement of vanishing of the matter SET trace, in connection with local scale symmetry (LSS) in curved spacetime, was derived in \cite{deser-1970}, in the case of a scalar field non-minimally coupled to gravity.\footnote{The vanishing of the matter SET trace as a requirement for conformal symmetry to hold \cite{ccj_ann_phys_1970} was primarily investigated within the HEP framework, where gravity is ignored, so the background spacetime is flat space with the Minkowski metric. Note, in this regard, that, within the HEP context, conformal symmetry refers to the 15-parameter conformal group that consists, in addition to the Poincar\'e group, of the 1-parameter group of dilatations and the 4-parameter special conformal group \cite{kastrup_1966, jackiw_1972}. These are spacetime transformations. In contrast, the conformal transformations that we consider in this paper are transformations of fields, including the spacetime metric and any matter fields that could be coupled to gravity, that do not affect either the spacetime points or the coordinates that label these points. That is, transformations \eqref{conf-t} are not spacetime diffeomorphisms. For that reason, in this document, to avoid confusion between conformal transformations \eqref{conf-t} and the HEP 15-parameter group of transformations, we will refer to the former mostly as Weyl transformations. In consequence, we say, for example, ``Weyl symmetry'' instead of ``conformal symmetry.''} The gravitational action of the theory investigated in this bibliographic reference, called here ``conformal general relativity'' (CGR), reads:\footnote{A Weyl symmetry-preserving quartic potential term $-\lambda\phi^4/12$, where $\lambda$ is a dimensionless constant, can be included within square brackets in \eqref{deser-action}. Here, for simplicity, we omit this term.}

\begin{align} S_\text{grav}=\frac{1}{2}\int d^4x\sqrt{-g}\left[\frac{1}{6}\,\phi^2R+(\der\phi)^2\right],\label{deser-action}\end{align} where $R$ is the curvature scalar, $\phi$ is the non-minimally coupled scalar field with mass dimensions, and we use the following shorthand notation: $(\der\phi)^2\equiv g^{\mu\nu}\der_\mu\phi\der_\nu\phi$. In \cite{deser-1970}, it is demonstrated that the form-invariance of \eqref{deser-action} plus a matter action piece $S_m=\int d^4x\sqrt{-g}\,L_m$ under LSTs \eqref{scale-t}, is possible only if the trace of the matter SET vanishes: $T^{(m)}\equiv g^{\mu\nu}T^{(m)}_{\mu\nu}=0$. Here, we shall demonstrate that the traceless condition $T^{(m)}=0$ is not a universal requirement for Weyl symmetry to materialize. The result of reference \cite{deser-1970}, that only matter fields with traceless SET can be coupled to gravity in CGR theory \eqref{deser-action}, is correct only if we assume that the masses of timelike fields are constant parameters that are not transformed by CTs \eqref{conf-t}, so the energy density of a perfect fluid transforms as: $\rho\rightarrow\hat\rho=\Omega^{-3}\rho$; that is, if we consider local scale transformations \eqref{scale-t} instead of Weyl transformations \eqref{units-t}. Under this assumption, the Lagrangian density of the matter fields breaks LSS. In contrast, if we assume that the masses of timelike fields are point-dependent fields that, under the CTs \eqref{conf-t}, transform as $m\rightarrow\hat m=\Omega^{-1}m$, and consequently the energy density of the perfect fluid transforms as $\rho\rightarrow\hat\rho=\Omega^{-4}\rho$, the Lagrangian density of the fundamental matter fields, as well as the Lagrangian density of perfect fluids, are form-invariant under the WTs \eqref{units-t}:

\begin{align} \int d^4x\sqrt{-g}\,L_m(\chi_i,g_{\mu\nu})=\int d^4x\sqrt{-\hat g}\,L_m(\hat\chi_i,\hat g_{\mu\nu}),\label{quiros}\end{align} and $\int d^4x\sqrt{-g}\,\rho=\int d^4x\sqrt{-\hat g}\,\hat\rho$, respectively. 


Our primary hypothesis in this work is that Dicke's principle holds in nature. That is, the laws of gravity are form-invariant under Weyl transformations \eqref{units-t}. In consequence, we consider that the masses of the fundamental timelike matter fields $m_i$ are point-dependent fields \cite{bekenstein_1980, hobson-prd-2020, hobson-epjc-2022, casas_1992}: $m_i(x)=\kappa_i\,\phi(x)$, where $\kappa_i$ are dimensionless constants.\footnote{This choice is dictated by the fact that, since the scalar field $\phi$ has mass dimensions, under CTs the mass parameter transforms as $\phi$ does; that is, $m\propto\phi$. In the bibliography, one can find other motivations for this choice (see, for instance, \cite{hobson-prd-2020, hobson-epjc-2022, casas_1992}).} These transform as in \eqref{units-t} under Weyl transformations, while the energy density of a perfect fluid transforms as $\rho\rightarrow\Omega^{-4}\rho$. In these cases, since the Lagrangian density of matter is form-invariant under WTs, the Ward identity associated with Weyl invariance leads to the following nonvanishing variational derivative:\footnote{In contrast, if the mass parameter $m$ is an untransformed constant, the Lagrangian density of the matter fields is not form-invariant under CTs \eqref{conf-t}, the Ward identity does not hold: $\delta{\cal L}_m/\delta\phi=0$.} $\delta{\cal L}_m/\delta\phi=\sqrt{-g}\,T^{(m)}/\phi$, which provides the term with the trace of the stress-energy tensor of matter in the EOM of the scalar field $\phi$. This allows any matter fields: timelike and null, to couple to gravity.


In view of Dicke's principle and subsequent form-invariance of the matter action under Weyl transformations \eqref{units-t}, it makes sense to investigate a Weyl-symmetric gravitational theory, whose simplest example could be based on the gravitational action \eqref{deser-action}. Here, we will investigate the phenomenological and physical consequences of the resulting Weyl-invariant theory of gravity, namely, the CGR theory. In particular, the possibility of finding an alternative explanation for two major cosmological puzzles: dark matter (DM) and dark energy (DE), as a consequence of Weyl symmetry, will be explored. The possibility that this could be an actual symmetry of the classical gravitational laws, which is contrary to the mainstream thinking about conformal symmetry --fuelled by the confusion of local scale transformations with Weyl transformations--, represents an unexplored alternative to solving the cosmological puzzles. 


We have organized the present document as follows. In Section \ref{sect-hist}, we set up the notation and briefly review the most important aspects of the work of reference \cite{deser-1970}, where the invariance of CGR under LSTs \eqref{scale-t} is considered. This is required for further comparison with the results of Section \ref{sect-var}, where the equations of motion (EOM) are derived from the overall CGR action under the assumptions of the present work, above all form-invariance of the matter action under WTs \eqref{units-t}. In this Section, we also derive some physical consequences of the CGR theory, in particular, the fifth force required by form-invariance under WTs, which does not interact with radiation, thus acting as a "dark force" with the potential to explain the ``dark sector'' of the present cosmological paradigm. In Section \ref{sect-phenom}, we discuss the two possible approaches to CTs in the configuration space \cite{quiros_prd_2025}: i) passive and ii) active approaches. We shall show that only the active approach is suitable for the investigation of the physical and phenomenological consequences of the Weyl symmetry. In Section \ref{sect-many-w}, we discuss gauge freedom within CGR, in connection with Weyl symmetry, under the active approach to WTs. We draw the similitude of the picture arising from gauge freedom within the CGR theory and the ``many-worlds'' interpretation of quantum mechanics \cite{everett_rmp_1957, wheeler, dewitt_phys_rev_1967, dewitt, barvinsky_cqg_1990, omnes_rmp_1992, tegmark_1998, garriga_prd_2001, zurek_rmp_2003, tegmark_nature_2007, quiros_prd_2023}. In Sections \ref{sect-dmat} and \ref{sect-cosmo}, we expose some important cosmological consequences of the present setup, including the possible explanation of the galactic rotation curves in some gauge of the CGR theory, as well as some encouraging aspects of Weyl-invariant cosmology. Possible observational consequences of CGR theory related to frequency redshift measurements are analyzed in Section \ref{sect-observ}, where a natural explanation of the measured redshifts of type Ia supernovae (SNIa) is provided without requiring dark energy. A discussion of quantum-mechanical removal of spacetime singularities in the present many-worlds framework is given in Section \ref{sect-sing}. This is followed by brief concluding remarks of the present work in Section \ref{sect-discu}. 



\section{Conformal general relativity and local scale symmetry}
\label{sect-hist}


The gravitational action of CGR theory \eqref{deser-action} is not only invariant but also form-invariant under the following conformal transformations of the gravitational fields:

\begin{align} g_{\mu\nu}\rightarrow\hat g_{\mu\nu}=\Omega^2g_{\mu\nu},\;\phi\rightarrow\hat\phi=\Omega^{-1}\phi.\label{gauge-t}\end{align} 

Independent variations of the CGR gravitational Lagrangian density 

\begin{align} {\cal L}_\text{grav}=\frac{1}{2}\sqrt{-g}\left[\frac{1}{6}\phi^2R+(\der\phi)^2\right],\label{deser-lag}\end{align} with respect to the metric and to the scalar field, neglecting divergence terms, yield \cite{deser-1970}:

\begin{align} \frac{\delta{\cal L}_\text{grav}}{\delta g^{\mu\nu}}=\sqrt{-g}\,\frac{\phi^2}{12}\,{\cal E}_{\mu\nu},\;\frac{\delta{\cal L}_\text{grav}}{\delta\phi}=\sqrt{-g}\left(\frac{1}{6}\,\phi R-\nabla^2\phi\right),\label{var-grav-lag}\end{align} where $G_{\mu\nu}\equiv R_{\mu\nu}-g_{\mu\nu}R/2$ is the Einstein tensor and

\begin{align} {\cal E}_{\mu\nu}\equiv G_{\mu\nu}-\frac{6}{\phi^2}\,\Theta^{(\phi)}_{\mu\nu},\;\Theta^{(\phi)}_{\mu\nu}\equiv\frac{1}{6}\left(\nabla_\mu\nabla_\nu-g_{\mu\nu}\nabla^2\right)\phi^2-T^{(\phi)}_{\mu\nu},\label{e-def}\end{align} where we use the shorthand notation $\nabla^2\equiv g^{\mu\nu}\nabla_\mu\nabla_\nu$. The modified SET of the scalar field $\Theta^{(\phi)}_{\mu\nu}$ \cite{ccj_ann_phys_1970}, has finite elements to all orders in the perturbation parameter and defines the same four-momentum and Lorentz generators as the conventional tensor:\footnote{Note that in the expression for $\Theta^{(\phi)}_{\mu\nu}$ the standard stress-energy tensor of the scalar field $T^{(\phi)}_{\mu\nu}$, has the wrong sign. This is because, in the gravitational Lagrangian density \eqref{deser-lag}, the kinetic energy density of the scalar field appears with the wrong sign, which is a requirement for the CGR gravitational Lagrangian density \eqref{deser-lag} to be invariant under CTs \eqref{gauge-t}. This has no physical consequences, since $\phi$ is a gauge field that does not obey a specific EOM.}

\begin{align} T^{(\phi)}_{\mu\nu}\equiv\der_\mu\phi\der_\nu\phi-\frac{1}{2}g_{\mu\nu}(\der\phi)^2.\label{phi-set}\end{align} If we take into account the trace of the tensor ${\cal E}_{\mu\nu}$: ${\cal E}\equiv g^{\mu\nu}{\cal E}_{\mu\nu}=-R+6\nabla^2\phi/\phi$, the equations in \eqref{var-grav-lag} can be written in the following form:

\begin{align} g^{\mu\nu}\frac{\delta{\cal L}_\text{grav}}{\delta g^{\mu\nu}}=\sqrt{-g}\frac{\phi^2}{12}\,{\cal E}=-\sqrt{-g}\frac{\phi}{2}\left(\frac{1}{6}\,\phi R-\nabla^2\phi\right),\;\frac{\phi}{2}\frac{\delta{\cal L}_\text{grav}}{\delta\phi}=\sqrt{-g}\frac{\phi}{2}\left(\frac{1}{6}\,\phi R-\nabla^2\phi\right),\nonumber\end{align} respectively. Therefore, the following Ward identity is obtained:

\begin{align} g^{\mu\nu}\frac{\delta{\cal L}_\text{grav}}{\delta g^{\mu\nu}}=-\frac{\phi}{2}\frac{\delta{\cal L}_\text{grav}}{\delta\phi}.\label{ward-0}\end{align} This identity is a consequence of the invariance of the CGR gravitational action \eqref{deser-action} under CTs \eqref{gauge-t}. Actually, if this action is form-invariant under infinitesimal CTs: 

\begin{align} \delta g_{\mu\nu}=2\theta\,g_{\mu\nu}\,\left(\delta g^{\mu\nu}=-2\theta\,g^{\mu\nu},\;\delta\sqrt{-g}=4\theta\sqrt{-g}\right),\;\delta\phi=-\theta\,\phi,\label{inf-gauge-t}\end{align} where for convenience (and only temporarily) we rescaled the conformal factor: $\Omega=e^\theta$, then $S_\text{grav}\rightarrow S_\text{grav}+\delta S_\text{grav}$, where $\delta S_\text{grav}=0$; so that, under \eqref{inf-gauge-t}, $S_\text{grav}\rightarrow S_\text{grav}$. We have;

\begin{align} \delta S_\text{grav}=\frac{1}{2}\int d^4x\left(\frac{\delta{\cal L}_\text{grav}}{\delta g^{\mu\nu}}\delta g^{\mu\nu}+\frac{\delta{\cal L}_\text{grav}}{\delta\phi}\delta\phi\right)=-\theta\int d^4x\left(g^{\mu\nu}\frac{\delta{\cal L}_\text{grav}}{\delta g^{\mu\nu}}+\frac{\phi}{2}\frac{\delta{\cal L}_\text{grav}}{\delta\phi}\right)=0,\nonumber\end{align} where in the last equality, we substituted $\delta g^{\mu\nu}$ and $\delta\phi$ from infinitesimal CTs \eqref{inf-gauge-t}. The Ward identity \eqref{ward-0} follows from the last equation.





Let us consider matter fields coupled to CGR; 

\begin{align} S_\text{tot}=\frac{1}{2}\int d^4x\sqrt{-g}\left[\frac{1}{6}\,\phi^2R+(\der\phi)^2\right]+\int d^4x\sqrt{-g}\,L_m(\chi_i,\der\chi_i,g_{\mu\nu}),\label{tot-action}\end{align} In \cite{deser-1970}, local scale transformations \eqref{scale-t} are considered rather than Weyl transformations \eqref{units-t}. Hence, it is assumed in the mentioned bibliographic reference that the CTs do not transform the masses; that is, the matter Lagrangian of time-like matter fields breaks LSS. In this case, the Ward identity for matter fields is not satisfied, so $\delta{\cal L}_m/\delta\phi=0$, because the Lagrangian density of matter ${\cal L}_m$ does not depend explicitly on $\phi$.


Variations of the total CGR action \eqref{tot-action} with respect to the metric $g_{\mu\nu}$ and with respect to $\phi$, yield

\begin{align} \frac{\delta{\cal L}_\text{tot}}{\delta g^{\mu\nu}}=\sqrt{-g}\,\frac{\phi^2}{12}\,\left[{\cal E}_{\mu\nu}-\frac{6}{\phi^2}\,T^{(m)}_{\mu\nu}\right],\;\frac{\delta{\cal L}_\text{tot}}{\delta\phi}=\sqrt{-g}\left(\frac{1}{6}\,\phi R-\nabla^2\phi\right),\label{var-tot-lag}\end{align} respectively, where we have omitted divergence terms and took into account that, $\delta{\cal L}_m/\delta\phi=0$ $\Rightarrow$ $\delta{\cal L}_\text{tot}/\delta\phi=\delta{\cal L}_\text{grav}/\delta\phi$. From \eqref{var-tot-lag} we get the following EOMs:

\begin{align} \frac{\delta{\cal L}_\text{tot}}{\delta g^{\mu\nu}}&=0\;\Rightarrow\;{\cal E}_{\mu\nu}=\frac{6}{\phi^2}\,T^{(m)}_{\mu\nu}\;\Rightarrow\;G_{\mu\nu}=\frac{6}{\phi^2}\left[\Theta^{(\phi)}_{\mu\nu}+T^{(m)}_{\mu\nu}\right],\label{einst-eom}\\
\frac{\delta{\cal L}_\text{tot}}{\delta\phi}&=0\;\Rightarrow\;\left(\nabla^2-\frac{1}{6}\,R\right)\phi=0.\label{kg-eom}\end{align} If we compare the trace of the Einstein EOM \eqref{einst-eom}:

\begin{align} -R+\frac{6}{\phi}\nabla^2\phi=\frac{6}{\phi^2}T^{(m)}\;\Rightarrow\;\left(\nabla^2-\frac{1}{6}\,R\right)\phi=\frac{1}{\phi}\,T^{(m)},\label{trace-eom}\end{align} with the Klein-Gordon (KG) type EOM \eqref{kg-eom}, we obtain the ``traceless condition:'' $T^{(m)}=0$. Therefore, only matter fields and fluids with vanishing SET trace: photons and radiation, in general, can be coupled to gravity in this local scale-invariant theory.\footnote{Non-relativistic matter fields that behave like radiation can also be coupled to CGR \cite{faria-a}.} 

This result has been the focus of studies on Weyl symmetry for decades, but not on the Weyl symmetry we are investigating in the present document. Recall that here, following \cite{bekenstein_1980}, we make a distinction between local scale transformations \eqref{scale-t} and Weyl transformations \eqref{units-t}, called by Dicke as transformations of units. Historically, LSTs have been the most studied Weyl-type transformations. In consequence, Weyl symmetry is usually identified with invariance under LSTs \eqref{scale-t}. Here, we associate Weyl symmetry with units' transformations \eqref{units-t}, which are popular in the study of scalar-tensor gravitational theories \cite{fujii_book, faraoni_book, flanagan_cqg_2004, faraoni_prd_2007, catena_prd_2007, chiba_2013, capozziello_prd_2013, jarv_2015, quiros_ijmpd_2019}.



\section{Conformal General Relativity and Weyl symmetry}
\label{sect-var}


The CGR theory \cite{deser-1970} whose total Lagrangian density is given by the following expression (see action \eqref{tot-action}):

\begin{align} {\cal L}_\text{tot}=\frac{\sqrt{-g}}{2}\left[\frac{1}{6}\,\phi^2R+(\der\phi)^2\right]+{\cal L}_m(\chi_i,\der\chi_i,g_{\mu\nu}),\label{tot-lag}\end{align} is form-invariant under the following Weyl transformations:

\begin{align} g_{\mu\nu}\rightarrow\hat g_{\mu\nu}=\Omega^2g_{\mu\nu},\;\phi\rightarrow\hat\phi=\Omega^{-1}\phi,\;\chi_i\rightarrow\hat\chi_i=\Omega^{w_{\chi_i}}\chi_i,\;m\rightarrow\hat m=\Omega^{-1}m,\label{weyl-t}\end{align} where it is assumed that the masses of timelike fields are themselves point-dependent fields that transform as $m\rightarrow\hat m=\Omega^{-1}m$, and, consequently, the energy density of perfect fluids transforms as $\rho\rightarrow\hat\rho=\Omega^{-4}\rho$.

Let us demonstrate that, under the above assumption, it is not necessary to make any changes in \eqref{tot-action}/\eqref{tot-lag} to get a consistent coupling of any matter fields $\chi_i$ to Weyl invariant gravity. That is, under these assumptions, the coupling of the matter fields in \eqref{tot-lag} preserves Weyl symmetry, regardless of whether $T^{(m)}$ vanishes. Our demonstration is based on the Weyl invariance of matter action, as demonstrated in \cite{quiros_prd_2025} for fundamental fields and perfect fluids. It has been demonstrated in the mentioned bibliographic reference that for fermions, gauge bosons, and perfect fluids, under the assumption that $m\rightarrow\hat m=\Omega^{-1}m$, the matter action \eqref{mat-action} is form-invariant under WTs. That is, under \eqref{weyl-t}, the following equality takes place:

\begin{align} {\cal L}_m(\chi_i,\der\chi_i,g_{\mu\nu})={\cal L}_m(\hat\chi_i,\hat\der\hat\chi_i,\hat g_{\mu\nu})\;\Leftrightarrow\;\sqrt{-g}\,L_m(\chi_i,\der\chi_i,g_{\mu\nu})=\sqrt{-\hat g}\,L_m(\hat\chi_i,\hat\der\hat\chi_i,\hat g_{\mu\nu}).\label{equiv}\end{align} From this equation it follows that if $g_{\mu\nu}$ is the physical metric with respect to the matter fields $\chi_i$; that is, the matter fields $\chi_i$ are minimally coupled to $g_{\mu\nu}$, the conformal metric $\hat g_{\mu\nu}$ is the physical metric with respect to the conformal matter fields $\hat\chi_i$ which are minimally coupled to $\hat g_{\mu\nu}$. In other words, if the metric $g_{\mu\nu}$ is the physical metric on some gauge, its conformal metric $\hat g_{\mu\nu}=\Omega^2g_{\mu\nu}$ is the physical metric on the conformal gauge. This result can also be stated in the following way: Weyl transformations \eqref{weyl-t} do not transform the minimal coupling of matter fields to the metric. Equation \eqref{equiv} plays an important role because it shows that the total Lagrangian density \eqref{tot-lag} is Weyl-invariant regardless of the type of matter coupled to gravity: fermions, gauge bosons, or perfect fluids.

Given that the Lagrangian density of matter fields ${\cal L}_m$ is form-invariant under WTs \eqref{weyl-t}, the variational derivative: $\delta{\cal L}_m/\delta\phi$ obeys the Ward identity; that is, $\delta{\cal L}_m/\delta\phi\neq 0$. This result, which we will immediately demonstrate, distinguishes our approach to the Weyl-symmetric CGR theory from the well-known result that the matter SET trace must vanish if local scale invariance is a symmetry of the CGR theory \cite{deser-1970}.

Let us apply the variational procedure under the assumption that the action of the matter degrees of freedom coupled to gravity: fermions, gauge bosons, and perfect fluids, is form-invariant under WTs \eqref{weyl-t}. We consider the Lagrangian density of timelike matter fields: ${\cal L}_m={\cal L}_m(\chi_i,\der\chi_i,g_{\mu\nu},\phi)$, where due to nonvanishing mass $m=\kappa\,\phi$, the Lagrangian density ${\cal L}_m$ depends explicitly on the gauge field $\phi$. For perfect fluids ${\cal L}_\text{fluid}=-\sqrt{-g}\,\rho$, since $\rho=\rho(\phi)$, then the Lagrangian density ${\cal L}_\text{fluid}$ also depends explicitly on $\phi$. Due to Weyl form-invariance, under infinitesimal Weyl transformations \eqref{inf-gauge-t} plus $\delta\chi_i=w_{\chi_i}\theta\chi_i$ and $\delta m=-\theta m$, we have $\delta{\cal L}_m=0$, or $\delta{\cal L}_m=\nabla_\mu V^\mu$, where $V^\mu$ is some vector. In any case, $\delta S_m=\int d^4x\,\delta{\cal L}_m=0$, so that under \eqref{inf-gauge-t}, $S_m\rightarrow S_m+\delta S_m=S_m$; that is, the matter action is invariant under infinitesimal Weyl transformations

\begin{align} \delta g_{\mu\nu}=2\theta\,g_{\mu\nu}\,\left(\delta g^{\mu\nu}=-2\theta\,g^{\mu\nu},\;\delta\sqrt{-g}=4\theta\sqrt{-g}\right),\;\delta\phi=-\theta\,\phi,\;\delta\chi_i=w_{\chi_i}\theta\chi_i,\;\delta m=-\theta m.\label{inf-gauge-t'}\end{align} If we take into account these infinitesimal Weyl transformations, for the variation of the Lagrangian density of matter, we have

\begin{align} \delta{\cal L}_m=\frac{\delta{\cal L}_m}{\delta g^{\mu\nu}}\delta g^{\mu\nu}+\frac{\delta{\cal L}_m}{\delta\phi}\delta\phi=0\;\Rightarrow\;-2\theta\left(g^{\mu\nu}\frac{\delta{\cal L}_m}{\delta g^{\mu\nu}}+\frac{\phi}{2}\frac{\delta{\cal L}_m}{\delta\phi}\right)=0,\label{var-lmat}\end{align} where, for simplicity, we have assumed that the EOMs of the matter fields are satisfied,\footnote{For most of classical gravitational problems, it is assumed that the matter fields are a priori given.} that is,

\begin{align} \frac{\delta{\cal L}_m}{\delta\chi_i}=0\;\Rightarrow\;\frac{\der{\cal L}_m}{\der\chi_i}-\nabla_\mu\left[\frac{\der{\cal L}_m}{\der(\der_\mu\chi_i)}\right]=0.\label{mat-eom}\end{align} From the equation on the right of \eqref{var-lmat}, it follows the Ward identity:

\begin{align} g^{\mu\nu}\frac{\delta{\cal L}_m}{\delta g^{\mu\nu}}=-\frac{\phi}{2}\frac{\delta{\cal L}_m}{\delta\phi}\;\Rightarrow\;\frac{\delta{\cal L}_m}{\delta\phi}=\frac{\sqrt{-g}}{\phi}\,T^{(m)},\label{mat-ward-id}\end{align} where, the equation on the right is obtained by substituting the definition of the matter SET \eqref{mat-set} in the Ward identity.


Let us now apply the variational principle to derive the equations of motion. We have

\begin{align} \delta S_\text{tot}&=\int d^4x\left[\left(\frac{\delta{\cal L}_\text{grav}}{\delta g^{\mu\nu}}+\frac{\delta{\cal L}_m}{\delta g^{\mu\nu}}\right)\delta g^{\mu\nu}+\left(\frac{\delta{\cal L}_\text{grav}}{\delta\phi}+\frac{\delta{\cal L}_m}{\delta\phi}\right)\delta\phi\right]\nonumber\\
&=\int d^4x\sqrt{-g}\left\{\frac{\phi^2}{12}\left[{\cal E}_{\mu\nu}-\frac{6}{\phi^2}T^{(m)}_{\mu\nu}\right]\delta g^{\mu\nu}+\left[-\nabla^2\phi+\frac{\phi}{6}R+\frac{1}{\phi}T^{(m)}\right]\delta\phi\right\}=0,\label{x}\end{align} where in the second line of \eqref{x}, in the factor multiplying $\delta\phi$ (expression within square brackets), we have taken into account the Ward identity \eqref{mat-ward-id}. The variational principle leads to the following EOMs:

\begin{align} {\cal E}_{\mu\nu}=\frac{6}{\phi^2}T^{(m)}_{\mu\nu},\label{einst-eom'}\\
\nabla^2\phi-\frac{\phi}{6}R=\frac{1}{\phi}T^{(m)}.\label{correct-kg-eom}\end{align} 

This is an outstanding result because the EOM \eqref{einst-eom'}, \eqref{correct-kg-eom}, are form-invariant under Weyl transformations, without requiring that the trace of the matter SET vanish. Note that the KG-EOM \eqref{correct-kg-eom} is not an independent equation, as it coincides with the trace of Einstein-type EOM \eqref{einst-eom'}. That is, the scalar field $\phi$ is not a dynamical field, as it does not satisfy an independent EOM. 



\subsection{The ``dark force''}


The independent EOM derived from \eqref{tot-lag}, is the Einstein-type equation \eqref{einst-eom'}. There is no independent EOM that the gauge field $\phi$ must obey, since the KG-type equation \eqref{correct-kg-eom}, which is derived from \eqref{tot-lag} by varying with respect to $\phi$, coincides with the trace of \eqref{einst-eom'}. The continuity equation provides another independent equation. If we take the divergence of \eqref{einst-eom'}, and use the second Bianchi identity $\nabla^\lambda G_{\lambda\mu}=0$, and the identity $(\nabla_\lambda\nabla_\mu-\nabla_\mu\nabla_\lambda)\nabla^\lambda\phi^2=R_{\mu\lambda}\nabla^\lambda\phi^2$, we obtain

\begin{align} \nabla^\lambda(\phi^2{\cal E}_{\lambda\mu})=6\der_\mu\phi\left(\nabla^2-\frac{1}{6}\,R\right)\phi=6\nabla^\lambda T^{(m)}_{\lambda\mu}.\nonumber\end{align} Then, if we substitute \eqref{correct-kg-eom} in the latter equation, we get the following inhomogeneous continuity equation:

\begin{align} \nabla^\lambda T^{(m)}_{\lambda\mu}=\frac{\der_\mu\phi}{\phi}\,T^{(m)}.\label{cont-eq}\end{align} 

The inhomogeneous term on the right-hand side (RHS) of \eqref{cont-eq} is required for the above equation to be form-invariant under WTs \eqref{weyl-t}. This term represents a fifth force, which can be related to the point-dependent property of the mass parameter in the present theory: $m(x)=\kappa\,\phi(x)$ \cite{bekenstein_1980, hobson-prd-2020, hobson-epjc-2022, casas_1992}. While \eqref{cont-eq} is satisfied by matter fluxes of stress and energy, the equivalent EOM for timelike, point-like fields, is the following EOM:

\begin{align} \frac{d^2x^\alpha}{ds^2}+\left\{^\alpha_{\mu\nu}\right\}\frac{dx^\mu}{ds}\frac{dx^\nu}{ds}=\frac{\der_\mu\phi}{\phi}\,h^{\mu\alpha},\label{timelike-eom}\end{align} where the orthogonal projection tensor is given by

\begin{align} h^{\mu\nu}\equiv g^{\mu\nu}-\frac{dx^\mu}{ds}\frac{dx^\nu}{ds}=g^{\mu\nu}+u^\mu u^\nu,\label{h-mn}\end{align} with the fourth velocity vector $u^\mu=dx^\mu/d\tau$ ($d\tau^2=-ds^2$) satisfying the following conditions: $g_{\mu\nu}u^\mu u^\nu=-1$ and $h_{\mu\lambda}u^\lambda=0$. Fields with vanishing mass, like photons and radiation fields in general, obey an actual conservation equation, $\nabla^\lambda T^\text{rad}_{\lambda\mu}=0,$ since their SET trace vanishes, $T^\text{rad}=g^{\mu\nu}T^\text{rad}_{\mu\nu}=0$. This means that these fields follow null geodesics of Riemann space $V_4$:

\begin{align} \frac{dk^\mu}{d\xi}+\left\{^\mu_{\nu\sigma}\right\}k^\nu k^\sigma=0,\label{0-geod}\end{align} where $k^\mu\equiv dx^\mu/d\xi$ is the wave vector ($k_\mu k^\mu=0$) and $\xi$ is an affine parameter along null-geodesic.\footnote{Note that the null-geodesic equation \eqref{0-geod} is form-invariant under CTs. Actually, under \eqref{conf-t} the wave vector transforms like $k^\mu\rightarrow\Omega^{-2}k^\mu$ $\Rightarrow$ $d\xi\rightarrow\Omega^2d\xi$, while: $$\frac{dk^\mu}{d\xi}\rightarrow\Omega^{-4}\left[\frac{dk^\mu}{d\xi}-2k^\mu\frac{d\ln\Omega}{d\lambda}\right],\;\{^\alpha_{\lambda\nu}\}k^\lambda k^\nu\rightarrow\Omega^{-4}\left[\{^\mu_{\lambda\nu}\}k^\lambda k^\nu+2k^\mu\frac{d\ln\Omega}{d\lambda}\right],$$ so that \eqref{0-geod} is not transformed by \eqref{conf-t}.} Therefore, the fifth force $f^\alpha=h^{\alpha\lambda}\der_\lambda\phi/\phi$ is orthogonal to the four-velocity: $u^\lambda f_\lambda=0$, and it acts on time-like fields exclusively. The fifth force does not interact with null-fields nor with radiation. For that reason, we call it ``dark force.'' This opens up the possibility that this extra force could explain, in principle, the dark energy and dark matter components of the cosmic fluid.


The assumption that $m=m(\phi)$ can have consequences for the standard model of particles (SMP). Actually, since timelike fields acquire masses through Yukawa interactions with the Higgs field $H$ after $SU(2)\times U(1)$ symmetry breaking, it is required to modify the SMP to get point-dependent masses. That is, the SMP must be Weyl-invariant, and the $SU(2)\times U(1)$ symmetry breaking must not spoil Weyl symmetry. Therefore, the Higgs Lagrangian density must be appropriately modified \cite{bars-2014, mohamedi-2024}. In particular, the mass parameter $v$ in the Higgs potential, $V\propto(H^\dag H-v^2)^2$, must be replaced by a field: $v\rightarrow v_0\phi$, where $v_0$ is some dimensionless parameter \cite{bars-2014}. After the mentioned modification, the timelike fields acquire point-dependent masses through the Yukawa interaction. For example, for a fermion field, the Yukawa coupling is given by $\kappa_f\bar\psi\phi\psi=\bar\psi\,m_f\,\psi$ $\Rightarrow$ $m_f=\kappa_f\,\phi$.

Notice that, as long as $\phi$ is the gravitational coupling in CGR theory, that is, $\phi\sim M_\text{pl}$, a mass scale issue could arise. However, we will never be able to measure the $\phi$-dependent mass, $m(x)=\kappa\,\phi(x)$, in local experiments, because measuring sticks and clocks are made of atoms (matter) that share the same universal $\phi$-dependence, so the mass unit also varies: $u_m(x)=u_0\,\phi(x)$. In experiments, we measure the ratio \cite{faraoni_prd_2007} 

\begin{align} \frac{m}{u_m}=\frac{\kappa}{u_0},\nonumber\end{align} that does not depend on the spacetime point. Hence, even if we assume $\phi_0\sim M_\text{pl}$, the value $\phi_0$ is not what is measured in experiments. In other words, the gigantic mass $m=\kappa\,\phi_0$ is compensated for by an equally gigantic mass unit $u_m=u_0\,\phi_0$, so there is no mass scale problem associated with point-dependent masses. The point-dependence of masses (and the related fifth force) can be observed only in ``nonlocal'' experiments. For example, in experiments that imply emission and absorption of photons coming from distant spacetime points, where a photon emitted by some atom located at some point $(t,\vec{x})$ is absorbed by an identical atom at the origin $(0,\vec{0})$. Since the mass of electrons is point-dependent, so is the energy of the emitted/absorbed photon. On the other hand, photon paths are not sensitive to WTs (the Riemannian null-geodesics are already Weyl-invariant). This mismatch makes it possible to reveal the point-dependence of the masses, but adequate experiments must be designed (see Section \ref{sect-observ} for further discussion of this subject).



\section{CGR: Physics beyond GR}
\label{sect-phenom}


It has been demonstrated in \cite{hobson-epjc-2022, woodard-1986} that there is no physics beyond GR in CGR theory.\footnote{In \cite{woodard-1986} CGR is named ``conformal scalar-metric theory'' in contrast to GR, named ``purely metric theory.''} While in \cite{woodard-1986}, it is shown that any ``conformal scalar-metric theory'' is gauge equivalent to the corresponding ``pure metric theory'', in \cite{hobson-epjc-2022}, it is argued that if one works in terms of
quantities that can be physically measured (in particular, the metric to which matter fields are minimally coupled), conformal Schwarzschild-de Sitter (SdS) metric does not lead to any different result than SdS metric; that is, ``in any conformal frame, the trajectories followed by ordinary matter particles are merely the timelike geodesics of the SdS metric.'' 

Here we shall show that the above-cited result, although correct, is not universal. It represents only one of two possible approaches to WTs \eqref{weyl-t}: i) passive Weyl transformations (PWTs), and ii) active Weyl transformations (AWTs) \cite{quiros_prd_2025, quiros-arxiv-2025}. Let us consider the configuration space; that is, an abstract space of fields ${\cal M}_f$ where the metric $g_{\mu\nu}$, the gauge field $\phi$, the matter fields $\chi_i$ coupled to gravity, and the masses $m$, are ``generalized coordinates.'' Each ``point'' $P:(\phi,g_{\mu\nu},\chi_i,m)$ in ${\cal M}_f$ represents a ``global gravitational state'' (GGS).\footnote{Here, by a global gravitational state we understand full knowledge of the fields at every spacetime point: $\phi=\phi(t,\vec{x})$, $g_{\mu\nu}=g_{\mu\nu}(t,\vec{x})$, $\chi_i=\chi_i(t,\vec{x})$ and $m=m(t,\vec{x})$. Hence, with each solution of the EOM of the CGR theory, one can associate a GGS. In Section \ref{sect-many-w}, we shall see that it is more appropriate to associate different points in the configuration space with different gauges rather than with different GGSs.} Passive Weyl transformations 

\begin{align} g_{\mu\nu}\rightarrow g'_{\mu\nu}=\Omega^2g_{\mu\nu},\;\phi\rightarrow\phi'=\Omega^{-1}\phi,\;\chi_i\rightarrow\chi'_i=\Omega^{w_{\chi_i}}\chi_i,\;m\rightarrow m'=\Omega^{-1}m,\label{pweyl-t}\end{align} may be viewed as a ``rotation'' of the coordinate system $R(g_{\mu\nu},\phi,\chi_i,m)$ $\rightarrow$ $R'(g'_{\mu\nu},\phi',\chi'_i,m')$ (see Figure 1 of \cite{quiros_prd_2025} or of \cite{quiros-arxiv-2025} where, for simplicity, the vacuum case is considered). The physically meaningful quantities are the invariant metric $\mathfrak{g}_{\mu\nu}$, matter fields $\Psi_i$, and mass $\mathfrak{m}$:

\begin{align} \mathfrak{g}_{\mu\nu}=\left(\frac{\phi}{\sqrt{6}M_\text{pl}}\right)^2\,g_{\mu\nu},\;\Psi_i=\left(\frac{\phi}{\sqrt{6}M_\text{pl}}\right)^{w_{\chi_i}}\chi_i,\;\mathfrak{m}=\left(\frac{\phi}{\sqrt{6}M_\text{pl}}\right)^{-1}m,\label{inv-q}\end{align} respectively. While $\mathfrak{g}_{\mu\nu}$, $\Psi_i$, and $\mathfrak{m}$ are the physical (thus measurable) quantities according to PWTs, the fields $g_{\mu\nu}$, $\phi$, $\chi_i$, and $m$, which are transformed according to \eqref{pweyl-t} are just ``auxiliary fields.'' The global gravitational state is given by $\mathfrak{G}_\mathfrak{g}:(\mathfrak{g}_{\mu\nu},\Psi_i,\mathfrak{m})$. Hence, $R$ and $R'$ are two different ``coordinate representations'' of the same point $P:(\mathfrak{g}_{\mu\nu},\Psi_i,\mathfrak{m})\in{\cal M}_f$; that is, of the same GGS.

It has been demonstrated in \cite{quiros_prd_2025} that the Lagrangian density of matter fields (this includes both fundamental matter fields and perfect fluids) is form-invariant under WTs \eqref{pweyl-t}. That is, the relationship \eqref{equiv} takes place. Similarly, it can be shown that the following equation \cite{quiros-arxiv-2025}:

\begin{align} {\cal L}_m(\chi_i,\der\chi_i,g_{\mu\nu})={\cal L}_m(\Psi_i,\mathfrak{D}\Psi_i,\mathfrak{g}_{\mu\nu}),\label{equiv'}\end{align} is also satisfied, where $\mathfrak{D}$ is the derivative operator defined in terms of the physical metric $\mathfrak{g}_{\mu\nu}$. This equality relates the Lagrangian density of the auxiliary matter fields $\chi_i$ to that of the physical matter fields $\Psi_i$. In \cite{quiros-arxiv-2025}, equality \eqref{equiv'} is shown for the Lagrangian density of the Dirac fermion and for that of perfect fluids.\footnote{It is straightforward to show that, for the Lagrangian density of the Dirac fermion: $\sqrt{-g}\,\bar\psi\left(i\cancel{\cal D}+m\right)\psi=\sqrt{-\mathfrak{g}}\,\bar\Psi\left(i\cancel{\mathfrak{D}}+\mathfrak{m}\right)\Psi$, where the gauge covariant derivative $\mathfrak{D}_\mu$ is defined in the same way as ${\cal D}_\mu$, but with the following replacements: $g_{\mu\nu}\rightarrow\mathfrak{g}_{\mu\nu}=(\phi/\sqrt{6}M_\text{pl})^2\,g_{\mu\nu}$, $e^a_\mu\rightarrow\mathfrak{e}^a_\mu=(\phi/\sqrt{6}M_\text{pl})\,e^a_\mu$, etc. Here, the conformal invariant fermion spinor and mass are given by $$\Psi=\left(\frac{\sqrt\phi}{M_\text{pl}}\right)^{-\frac{3}{2}}\psi,\;\mathfrak{m}=\left(\frac{M_\text{pl}}{\sqrt\phi}\right)\,m,$$ respectively. The equality of the Lagrangian density of the Dirac fermion in terms of the auxiliary fields with the one in terms of the conformal invariant fields implies that ${\cal L}_m(\psi,{\cal D}\psi,g_{\mu\nu})={\cal L}_m(\Psi,\mathfrak{D}\Psi,\mathfrak{g}_{\mu\nu})$, thus confirming \eqref{equiv'}. For perfect fluids, the demonstration of \eqref{equiv'} is also simple. The Lagrangian density for a perfect fluid with energy density $\rho$ can be written as ${\cal L}_\text{fluid}[\rho,g]=-\sqrt{-g}\,\rho$. If we define the Weyl invariant energy density $\mathfrak{r}:=(M_\text{pl}/\sqrt{6}\phi)^4\rho$, and take into account the definition of the Weyl invariant metric tensor \eqref{inv-q}, then, since $\sqrt{-\mathfrak{g}}=(\sqrt{6}\phi/M_\text{pl})^4\sqrt{-g}$, we get that $\sqrt{-\mathfrak{g}}\,\mathfrak{r}=\sqrt{g}\,\rho$; that is, ${\cal L}_\text{fluid}[\rho,g]={\cal L}_\text{fluid}[\mathfrak{r},\mathfrak{g}]$, which is \eqref{equiv'} for the case of perfect fluids.} For other fundamental fields, e. g., gauge bosons, the demonstration is similar.

In view of equality \eqref{equiv'}, the equation \eqref{tot-lag} can be written as

\begin{align} {\cal L}_\text{tot}=\frac{\sqrt{-g}}{2}\left[\frac{1}{6}\,\phi^2R+(\der\phi)^2\right]+{\cal L}_m(\Psi_i,\mathfrak{D}\Psi_i,\mathfrak{g}_{\mu\nu}).\label{tot-lag'}\end{align} The next step is to write the gravitational part of this Lagrangian density in terms of the physical metric, just as we have done for the matter Lagrangian density. We get

\begin{align} {\cal L}_\text{tot}=\sqrt{-\mathfrak{g}}\,\frac{M^2_\text{pl}}{2}\mathfrak{R}+{\cal L}_m(\Psi_i,\mathfrak{D}\Psi_i,\mathfrak{g}_{\mu\nu}),\label{tot-lag-pwt}\end{align} where $\mathfrak{R}$ is the curvature scalar defined in terms of the Weyl-invariant, thus physical, metric. Written in this covariant way in the configuration space, the total Lagrangian density coincides with that of GR. The difference between \eqref{tot-lag} and \eqref{tot-lag-pwt} according to the PWTs is that the latter Lagrangian density is written in terms of physical quantities. In contrast, the former one is written in terms of auxiliary fields without a direct physical meaning. That is, according to the passive approach to WTs, we can safely remove the auxiliary fields and the related Weyl symmetry (the WTs act precisely on the auxiliary fields) from the covariant formulation \eqref{tot-lag-pwt} of the CGR theory in the configuration space. Therefore, the passive approach to WTs is not suitable for searching for the physical and phenomenological consequences of Weyl symmetry. In this case, as demonstrated in \cite{woodard-1986} and confirmed in \cite{hobson-epjc-2022}, there is no physics beyond GR in CGR theory. However, as we have already commented, this is only one of two possible understandings of WTs \eqref{weyl-t}.


In contrast to passive transformations, AWTs represent real ``motion'' in the configuration space ${\cal M}_f$. That is, the ``coordinate system'' is held fixed, so $P:(g_{\mu\nu},\phi,\chi_i,m)$ and $\hat P:(\hat g_{\mu\nu},\hat\phi,\hat\chi_i,\hat m)$ are different points in ${\cal M}_f$. Hence, the Weyl transformations;

\begin{align} g_{\mu\nu}\rightarrow \hat g_{\mu\nu}=\Omega^2g_{\mu\nu},\;\phi\rightarrow\hat\phi=\Omega^{-1}\phi,\;\chi_i\rightarrow\hat\chi_i=\Omega^{w_{\chi_i}}\chi_i,\;m\rightarrow\hat m=\Omega^{-1}m,\label{aweyl-t}\end{align} relate two different GGSs, $\mathfrak{G}:(\phi,g_{\mu\nu},\chi_i,m)$, and $\hat{\mathfrak{G}}:(\hat\phi,\hat g_{\mu\nu},\hat\chi_i,\hat m)$ in the configuration space. In other words, the overall Lagrangian density \eqref{tot-lag}: ${\cal L}_\text{tot}=\sqrt{-g}\left[\phi^2R/6+(\der\phi)^2\right]/2+{\cal L}_m(\chi_i,\der\chi_i,g_{\mu\nu})$, is already written covariantly in terms of physical fields $g_{\mu\nu}$, $\phi$, $\chi_i$, and $m$. In this case, Weyl invariants, such as those defined in \eqref{inv-q}, represent no more than useful relationships between the physical fields in the two different global gravitational states:

\begin{align} \mathfrak{g}_{\mu\nu}&=\left(\frac{\phi}{\sqrt{6}M_\text{pl}}\right)^2\,g_{\mu\nu}=\left(\frac{\hat\phi}{\sqrt{6}M_\text{pl}}\right)^2\,\hat g_{\mu\nu}\;\Rightarrow\;g_{\mu\nu}=\left(\frac{\hat\phi}{\phi}\right)^2\,\hat g_{\mu\nu},\nonumber\\
\Psi_i&=\left(\frac{\phi}{\sqrt{6}M_\text{pl}}\right)^{w_{\chi_i}}\chi_i=\left(\frac{\hat\phi}{\sqrt{6}M_\text{pl}}\right)^{w_{\chi_i}}\hat\chi_i\;\Rightarrow\;\chi_i=\left(\frac{\hat\phi}{\phi}\right)^{w_{\chi_i}}\hat\chi_i,\nonumber\\\mathfrak{m}&=\left(\frac{\phi}{\sqrt{6}M_\text{pl}}\right)^{-1}m=\left(\frac{\hat\phi}{\sqrt{6}M_\text{pl}}\right)^{-1}\hat m\;\Rightarrow\;m=\left(\frac{\hat\phi}{\phi}\right)^{-1}\hat m.\label{rel}\end{align} 

From now on, when we talk about Weyl transformations, we mean the active approach.



\subsection{Cautionary note on gauge freedom}


The interpretation of gauge freedom that we have applied within the framework of CGR theory, as well as its physical meaning, has nothing in common with the usual interpretation of gauge freedom in standard gauge theories of fields. Consider, for example, the Lagrangian of a Maxwell field $A_\mu$ coupled to a Dirac fermion $\psi$: 

\begin{align} {\cal L}_\text{em}=-\frac{1}{4}\,F_{\mu\nu}F^{\mu\nu}+\bar\psi(i\cancel{D}-m)\psi,\label{em-lag}\end{align} where $F_{\mu\nu}\equiv\der_{[\mu}A_{\nu]}/2$ is the Maxwell field strength, $\psi$ is the Dirac spinor of the fermion field ($\bar\psi$ is its adjoint), $\gamma^\mu$ are the Dirac gamma matrices, and $\cancel{D}\psi\equiv\gamma^\mu D_\mu\psi=\gamma^\mu(\der_\mu\psi+iA_\mu\psi)$ is the gauge covariant derivative. In \eqref{em-lag} it is assumed that the background space is Minkowski space. This is the first clear difference with the present setup, because the conformal transformations \eqref{conf-t} act only on the metric of curved spaces. 

The Lagrangian \eqref{em-lag} is form-invariant under gauge transformations: 

\begin{align} A_\mu\rightarrow\hat A_\mu= A_\mu+\der_\mu\lambda(x),\;\psi\rightarrow\hat\psi=e^{-i\lambda(x)}\psi,\;\bar\psi\rightarrow\hat{\bar\psi}=\bar\psi\,e^{i\lambda(x)},\nonumber\end{align} where the free function $\lambda(x)$ is the gauge parameter. In this case, gauge invariance of \eqref{em-lag} means that the derived EOM, as well as the measured quantities; that is, the physically meaningful quantities: the Lorentz scalar type $\bar\psi\psi$, vector type $\bar\psi\gamma^\mu\psi$, tensor type $\bar\psi\gamma^\mu\gamma^\nu\psi$ quantities, and the time and spatial components of the gauge invariant field strength $F_{\mu\nu}$; the electric $E_i=F_{0i}$ and magnetic $B_i=-\epsilon_{ijk}F^{jk}/2$ fields ($\epsilon_{ijk}$ is the Levi-Civita tensor), respectively, are not affected by the gauge transformations; that is, $\bar\psi\psi=\hat{\bar\psi}\hat\psi$, $\bar\psi\gamma^\mu\psi=\hat{\bar\psi}\gamma^\mu\hat\psi$, ..., $E_i=\hat E_i$ and $B_i=\hat B_i$. This means that these transformations do not modify the state of the Maxwell field in interaction with the fermion. Note that, in the present case, the fields $A_\mu$, $\psi$, and $\bar\psi$, by themselves, do not represent measured quantities, so these are physically meaningless. Even if we try to associate different electromagnetic (EM) states with different sets of these fields: $\mathfrak{E}:(\psi,\bar\psi, A_\mu)$, $\hat{\mathfrak{E}}:(\hat\psi,\hat{\bar\psi}, \hat A_\mu)$, etc., these states share the same measured quantities so that they are physically undistinguishable. To all practical purposes, $\mathfrak{E}$ and $\hat{\mathfrak{E}}$ represent the same physical EM state. This is in contrast with the active interpretation of Weyl transformations of the physically meaningful quantities; that is, the metric, the scalar field, and the other fields coupled to gravity: $g_{\mu\nu}\rightarrow\hat g_{\mu\nu}=\Omega^2g_{\mu\nu}$, $\phi\rightarrow\hat\phi=\Omega^{-2}\phi$, ..., where $\{\phi(x),g_{\mu\nu}(x),...\}$ and $\{\hat\phi(x),\hat g_{\mu\nu}(x),...\}$ represent different global gravitational states, which are characterized by different measured quantities; that is, different curvature properties, proper time lapse, etc. 

In this document, we call gauge fixing the specific choice of the free scalar $\phi(x)$, as explained above, because, historically, Weyl called conformal transformations \eqref{conf-t} as ``gauge transformations''.



\section{Gauge freedom: The ``many-worlds'' interpretation}
\label{sect-many-w}


The freedom to choose any arbitrary function $\phi=\phi(x)$ is called ``gauge freedom''. Different gauges are fixed by different choices of the scalar $\phi=\phi(x)$. For example, the GR gauge is achieved by choosing $\hat\phi=\sqrt{6}\,M_\text{pl}=$ const.\footnote{We underline that once we make an specific gauge choice $\phi=\phi(x)$, the explicit Weyl invariance is lost. This is similar to the loss of explicit coordinate invariance after a specific choice of coordinates is made within the GR framework.} Under this choice, the total Lagrangian density \eqref{tot-lag} is written as;

\begin{align} {\cal L}_\text{tot}=\sqrt{-\hat g}\,\frac{M^2_\text{pl}}{2}\,\hat R+{\cal L}_m(\hat\chi_i,\hat\der\hat\chi_i,\hat g_{\mu\nu}),\label{tot-lag-transf}\end{align} while the Einstein EOM \eqref{einst-eom'} reads

\begin{align} \hat G_{\mu\nu}=\frac{1}{M^2_\text{pl}}\,\hat T^{(m)}_{\mu\nu},\label{gr-eom}\end{align} which is the standard GR Einstein EOM. The GR gauge is related to any other (arbitrary) gauge of \eqref{tot-lag} through the following equations (see \eqref{rel}):

\begin{align} g_{\mu\nu}\rightarrow\hat g_{\mu\nu}=\left(\frac{\phi}{\sqrt{6}M_\text{pl}}\right)^2g_{\mu\nu},\;\chi_i\rightarrow\hat\chi_i=\left(\frac{\phi}{\sqrt{6}M_\text{pl}}\right)^{w_{\chi_i}}\chi_i,\;m\rightarrow\hat m=\left(\frac{\sqrt{6}M_\text{pl}}{\phi}\right)\,m,\nonumber\end{align} where $\hat m$ is a constant mass parameter and we took into account $\phi\rightarrow\hat\phi=\Omega^{-1}\phi=\sqrt{6}M_\text{pl}$, from which we get $\Omega=\phi/\sqrt{6}M_\text{pl}$. Here, we also consider the transformation $T^{(m)}_{\mu\nu}\rightarrow\hat T^{(m)}_{\mu\nu}=\Omega^{-2}\,T^{(m)}_{\mu\nu}$ of the matter SET; that is, $\hat T^{(m)}_{\mu\nu}=(\phi/\sqrt{6}M_\text{pl})^{-2}\,T^{(m)}_{\mu\nu}$. Since GR is a gauge of Weyl-invariant theory \eqref{tot-lag}, we refer to the latter theory as CGR. 

Note that in an arbitrary gauge, where ${\cal L}_m={\cal L}_m(\chi_i,\der\chi_i,g_{\mu\nu})$, the matter fields $\chi_i$ are minimally coupled to the metric $g_{\mu\nu}$. Therefore, it is the metric that defines the proper time intervals; that is, it is the physical metric in that gauge \cite{brans-1988}. In the same way, the matter fields $\hat\chi_i$ are minimally coupled to the metric $\hat g_{\mu\nu}$, so it is the physical metric in the conformal gauge. Let us take, for example, the GR gauge $\mathfrak{G}_\text{gr}=\left\{\sqrt{6}M_\text{pl},\hat g_{\mu\nu}(x),\hat\chi_i,\hat m\right\}$, where $\hat g_{\mu\nu}=\hat g_{\mu\nu}(x)$ is any solution of the GR-EOM \eqref{gr-eom} ($\hat m=$ const. is the GR constant mass parameter), and any other arbitrary gauge $\mathfrak{G}=\{\phi(x),g_{\mu\nu}(x),\chi_i,m\}$, where $g_{\mu\nu}=g_{\mu\nu}(x)$ is a solution of the Einstein-type EOM \eqref{einst-eom'} once we have fixed the function $\phi=\phi(x)$ (it is assumed that the matter fields are given). These gauges represent potential descriptions of the conformal invariant gravitational interactions of matter according to CGR theory, where the respective solutions are related by

\begin{align} g_{\mu\nu}(x)=\left(\frac{\sqrt{6}M_\text{pl}}{\phi(x)}\right)^2\hat g_{\mu\nu}(x).\label{invert-gauge-t}\end{align} Hence, we do not need to solve the Einstein-type EOM \eqref{einst-eom'} once we make a specific choice for $\phi=\phi(x)$. We only need to know some GR solution $\hat g_{\mu\nu}(x)$, then, if we fix the function $\phi=\phi(x)$, the corresponding conformal metric, according to \eqref{invert-gauge-t}, is necessarily a solution of \eqref{einst-eom'}. 


The usefulness of this result is illustrated, for example, in a series of papers \cite{bambi-2017-a, bambi-2017-b, bambi-2017-c, bambi-2018-a, bambi-2018-b}, where Weyl invariance is used in several contexts, for example, to look for the properties of some astrophysical black holes, to investigate the black hole evaporation, as well as the singularity issues. Note that in these bibliographic references, each conformal solution is assumed to be physically meaningful. However, this is true only if we follow the active approach to Weyl transformations.\footnote{In contrast, if we follow the passive approach, the physical metric is the conformal invariant quantity $$\mathfrak{g}_{\mu\nu}=\left(\frac{\phi}{\sqrt{6}M_\text{pl}}\right)^2\,g_{\mu\nu}=\hat g_{\mu\nu}.$$ That is, the GR metric $\hat g_{\mu\nu}$ coincides with the physical metric. Therefore, only the GR solution $\hat g_{\mu\nu}(x)$ matters.} In this case, the GR gauge $\mathfrak{G}_\text{gr}$ and any other arbitrary gauge $\mathfrak{G}_k$; 

\begin{align} \mathfrak{G}_k=\{\phi_k(x),g^{(k)}_{\mu\nu}(x),...\},\;g^{(k)}_{\mu\nu}=\left(\frac{\sqrt{6}M_\text{pl}}{\phi_k}\right)^2\hat g_{\mu\nu},\;k=1,2,...,N-1,\label{gauge-def}\end{align} where $N\rightarrow\infty$ and the ellipsis within curled brackets stands for other fields coupled to gravity, belong in the conformal equivalence class (CEC) 

\begin{align} \mathfrak{C}_\text{conf}=\{\mathfrak{G}_\text{gr},\mathfrak{G}_1,\mathfrak{G}_2,...,\mathfrak{G}_{N-1}\},\label{c-class}\end{align} of CGR theory given by the Lagrangian density \eqref{tot-lag} and the derived EOM \eqref{einst-eom'}. As a result, these gauges correspond to different global gravitational states, each with distinct phenomenological consequences. The choice between one gauge and another is ultimately an experimental matter. The arising picture may be understood as a classical version of the many-worlds interpretation of quantum mechanics \cite{everett_rmp_1957, wheeler, dewitt_phys_rev_1967, dewitt, barvinsky_cqg_1990, omnes_rmp_1992, tegmark_1998, garriga_prd_2001, zurek_rmp_2003, tegmark_nature_2007, quiros_prd_2023} applied to the gravitational interactions of matter. However, in the present case, the many worlds picture arises not in spacetime but in the configuration space. 


Although we have treated the concepts of a gauge and a global gravitational state as almost the same, there is a distinction between them. To illustrate the difference, let us consider different solutions of the GR-EOM for different matter sources: ($\hat g^{(1)}_{\mu\nu}(x)$, $\chi_i^{(1)}$), ($\hat g^{(2)}_{\mu\nu}(x)$, $\chi_i^{(2)}$), ..., ($\hat g^{(N_s)}_{\mu\nu}(x)$, $\chi_i^{(N_s)}$), where $N_s$ is the number of different solutions. We can label the different solutions by the index ``$A$'' ($A=1,2,...,N_s$), such that, in compact form: ($\hat g^{(A)}_{\mu\nu}(x)$, $\chi_i^{(A)}$). Hence, in the GR gauge, we can have $N_s$ global gravitational states: $\mathfrak{G}(A|\text{gr})$, each one corresponding to a possible solution of the GR equations of motion. Any other gauge ``$k$'' in the same conformal equivalence class can be expressed as $\mathfrak{G}(A|k)$. Therefore, we now have one CEC associated with each solution or GGS:

\begin{align} \mathfrak{C}^A_\text{conf}=\{\mathfrak{G}(A|\text{gr}),\mathfrak{G}(A|1),\mathfrak{G}(A|2),...,\mathfrak{G}(A|N-1)\},\label{cec-a}\end{align} and the universal CEC: $\mathfrak{C}^\text{univ}_\text{conf}=\{\mathfrak{C}^1_\text{conf}, \mathfrak{C}^2_\text{conf}, ..., \mathfrak{C}^{N_s}_\text{conf}\}$.



\subsection{Vacuum solution: Conformal Schwarzschild space}


In a series of papers \cite{bambi-2017-a, bambi-2017-b, bambi-2017-c, bambi-2018-a, bambi-2018-b}, conformal invariance is used to investigate the properties of astrophysical black holes and to examine black hole evaporation and singularity issues. In these bibliographic references, each conformal solution is assumed to be physically meaningful; that is, the authors implicitly assume the active approach to WTs. Imagine, for example, that in the GR gauge we have the Schwarzschild GR vacuum solution:

\begin{align} d\hat s^2=\hat g_{\mu\nu}dx^\mu dx^\nu=-\left(1-\frac{2m}{r}\right)dt^2+\frac{dr^2}{\left(1-\frac{2m}{r}\right)}+r^2d\Omega^2,\label{schw-gr}\end{align} where $d\Omega^2\equiv d\theta^2+sin^2\theta\,d\vphi^2$. Then, according to \eqref{invert-gauge-t}, the conformal metric:

\begin{align} ds^2=g_{\mu\nu}dx^\mu dx^\nu=\left[\frac{\sqrt{6}M_\text{pl}}{\phi(r)}\right]^2\left[-\left(1-\frac{2m}{r}\right)dt^2+\frac{dr^2}{\left(1-\frac{2m}{r}\right)}+r^2d\Omega^2\right],\label{conf-schw-gr}\end{align} is also a solution of the vacuum EOM\footnote{The solutions \eqref{schw-gr} and their conformal \eqref{conf-schw-gr}, are valid in the absence of a cosmological term. Otherwise, we must replace \eqref{schw-gr} with the Schwarzschild-de Sitter solution instead.} 

\begin{align} {\cal E}_{\mu\nu}\equiv G_{\mu\nu}+\frac{6}{\phi^2}\left[\der_\mu\phi\der_\nu\phi-\frac{1}{2}g_{\mu\nu}(\der\phi)^2\right]-\frac{1}{\phi^2}\left(\nabla_\mu\nabla_\nu-g_{\mu\nu}\nabla^2\right)\phi^2=0.\label{vac-eom}\end{align} 

As an illustration of the meaning of gauge fixing in CGR within the AWTs framework, in the bibliographic reference \cite{bambi-2017-a}, the authors made the following gauge choice:\footnote{In this and other bibliographic references, the BD scalar field $\Phi=\phi^2/6$ is used instead of $\phi$, so certain minimal differences may appear in the resulting formulae with respect to the present document.}

\begin{align} \phi(r)=\frac{\sqrt{6}M_\text{pl}}{\sqrt{1+(L/r)^4}}=\frac{\sqrt{6}M_\text{pl}\,r^2}{\sqrt{r^4+L^4}},\label{bambi-phi}\end{align} where $L$ is some length scale. It is shown in \cite{bambi-2017-a} that the spacetime given by the metric \eqref{conf-schw-gr}, with $\phi(r)$ given by \eqref{bambi-phi}, is complete, so the resulting geometry represents a nonsingular black hole. The authors conclude that the singularity issue is an artifact of the GR gauge. According to \cite{bambi-2017-a}, another possible gauge choice, which is suitable for removing the spacetime singularity, could be

\begin{align} \phi(r)=\frac{\sqrt{6}M_\text{pl}}{1+(L/r)^2}=\frac{\sqrt{6}M_\text{pl}\,r^2}{r^2+L^2}.\label{bambi-other}\end{align} 

In these examples, the Schwarzschild vacuum solution $d\hat s^2$ given by \eqref{conf-schw-gr} and its conformal singularity-free solutions; $[\sqrt{6}M_\text{pl}/\phi(r)]^2\,d\hat s^2$, where $\phi(r)$ can be given by \eqref{bambi-phi} or \eqref{bambi-other} (or possibly by many other $r$-functions), give suitable descriptions of vacuum CGR \eqref{vac-eom}. The fact that global gravitational states with a spacetime singularity are in the same equivalence class as singularity-free spacetimes is very interesting because this might shed some light on the singularity issue, as done, for example, in \cite{bambi-2017-a, bambi-2017-b, bambi-2017-c, bambi-2018-a, bambi-2018-b}.\footnote{If instead of the AWTs we adopted the passive approach to WTs, the results in \cite{bambi-2017-a, bambi-2017-b, bambi-2017-c, bambi-2018-a, bambi-2018-b} would not have any physical significance. In fact, according to passive WTs, the physically meaningful metric $\mathfrak{g}_{\mu\nu}=[\phi(r)/\sqrt{6}M_\text{pl}]^2\,g_{\mu\nu}=\hat g_{\mu\nu}$, coincides with the GR vacuum metric. That is, only the Schwarzschild black hole solution \eqref{schw-gr} matters.} Yet, it is ultimately the experiment that decides which gauge is the one that better describes the gravitational phenomena. If it were, for example, the GR gauge, the one selected by experiment, then the spacetime singularity would still be there. Nevertheless, as we will show in Section \ref{sect-sing}, consideration of quantum-mechanical effects in the present framework may remove the classical spacetime singularity.



\section{Galactic rotation curves}
\label{sect-dmat}


The possibility that the observed behavior of test particles outside galaxies, which is usually explained with the help of dark matter, could be explained, alternatively, within the framework of so-called conformal invariant Weyl geometric quadratic gravity \cite{ghilen-2019}, was investigated in \cite{harko-prd-2023}. The static spherically symmetric solution found in \cite{harko-epjc-2022} of the EOM of the latter theory was one of three cases studied in \cite{harko-prd-2023}, showing a good fit to the galactic rotation curves. 
 
For the specific form of the Weyl vector with only a radial nonvanishing component, chosen in \cite{harko-prd-2023, harko-epjc-2022}, the resulting vacuum EOMs coincide exactly with \eqref{vac-eom} with the addition of a Weyl symmetry-preserving potential $\phi^4/24$ (in \cite{harko-prd-2023, harko-epjc-2022} the BD field $\Phi=\phi^2$ was used). Then, there is no doubt that the solution found in these publications is a vacuum solution of Weyl-invariant CGR theory \eqref{tot-lag} with a cosmological constant. As we shall see, the solution involves a specific function $\phi=\phi(r)$, so that it is in fact a gauge of vacuum CGR theory \eqref{vac-eom} with a cosmological constant. Let us discuss the mentioned solution that provides a good explanation of the galaxies' rotation curves.

In \cite{harko-prd-2023, harko-epjc-2022} the static spherically symmetric metric 

\begin{align} ds^2=-e^{\nu(r)}dt^2+e^{\lambda(r)}dr^2+r^2d\Omega^2,\label{sss-sol}\end{align} was assumed. The solution studied in \cite{harko-prd-2023} reads:

\begin{align} e^{\nu}=e^{-\lambda}=\alpha-2+\frac{(\alpha-3)C_2}{3r}+(\alpha-1)\frac{r}{C_2}+C_3r^2,\label{harko-sol}\end{align} which corresponds to the gauge choice:

\begin{align} \phi(r)=\frac{\sqrt{C_1}M_\text{pl}}{r+C_2}.\label{harko-gauge}\end{align} In these equations $C_1$, $C_2$ and $C_3$ are integration constants, while $\alpha\equiv 3C_3C^2_2+C_1/4$. It is not difficult to show that, under the coordinate transformation,

\begin{align} \tilde t=\sqrt\frac{C_1}{6C^2_2}\,t,\;\tilde r=\sqrt\frac{C_1}{6}\frac{r}{r+C_2},\nonumber\end{align} the above solution is written as conformal Schwarzschild-de Sitter (SdS) solution:\footnote{The study of conformal SdS geometry to explain the galactic rotation curves has been undertaken also in \cite{modesto-2020, modesto-2024}.}

\begin{align} ds^2=\left[\frac{\sqrt{6}M_\text{pl}}{\phi(\tilde r)}\right]^2d\hat s^2,\label{schw-ds}\end{align} where

\begin{align} d\hat s^2=-\left(1-\frac{2m}{\tilde r}-\frac{\Lambda_0}{3}\,\tilde r^2\right)\,d\tilde t^2+\left(1-\frac{2m}{\tilde r}-\frac{\Lambda_0}{3}\,\tilde r^2\right)^{-1}d\tilde r^2+\tilde r^2d\Omega^2,\label{schw-ds-gr}\end{align} is the GR-SdS solution and the choice of the scalar field \eqref{harko-gauge} in the new coordinates reads as

\begin{align} \phi(\tilde r)=\frac{\sqrt{6}M_\text{pl}}{C_2}\left(\sqrt\frac{C_1}{6}-\tilde r\right).\nonumber\end{align} In these equations, we have defined the following constants: $2m\equiv\sqrt{C_1/6}(1-C_3C^2_2-C_1/12)$ and $\Lambda_0\equiv C_1/4$. 

This is an expected result, since \eqref{harko-sol} is a vacuum solution of CGR with a quartic potential $-\lambda\phi^4/12$ within square brackets in \eqref{tot-lag}, so the derived Einstein-like EOM reads as: ${\cal E}_{\mu\nu}+\lambda\phi^2g_{\mu\nu}/4=0$. In this case, the gauge choice $\phi=\sqrt{6}M_\text{pl}$ leads to GR plus a cosmological constant $\Lambda=3\lambda M^2_\text{pl}/2$. Hence, any solution \eqref{schw-ds} that is conformal to the Schwarzschild-de Sitter vacuum solution \eqref{schw-ds-gr} is also a solution of \eqref{vac-eom} plus a quartic potential term $\lambda\phi^2g_{\mu\nu}/4$.

According to active CTs, the GR Schwarzschild-de Sitter metric \eqref{schw-ds-gr}, which is not suitable for explaining the galactic rotation curves, and the conformal solution \eqref{schw-ds}, which can provide a good background for explaining the rotation curves, belong to the same CEC $\mathfrak{C}_\text{conf}$ \eqref{c-class}. Both are equally plausible solutions of vacuum EOM \eqref{vac-eom} (plus a quartic potential term) that represent different global gravitational states according to AWTs. Only experiment (in the present case, the observational data on galactic rotation curves) can choose a specific gauge among the infinity of conformally related gauges in the conformal class $\mathfrak{C}_\text{conf}$ of vacuum CGR theory plus a quartic potential term.

If we look at the above problem from the point of view of the passive approach to CTs, as done, for instance, in \cite{hobson-epjc-2022}, there are no physical consequences of Weyl symmetry. Actually, in this bibliographic reference, it is stated that if one works in terms of quantities that can be physically measured, within the framework of conformally invariant gravity theories defined on Riemannian spacetime and having the SdS metric as a solution in the Einstein gauge, in any conformal frame, the trajectories followed by matter particles are merely the timelike geodesics of the SdS metric. Hence, there is no frame dependence of physical predictions, and it is clear that the rotation curves have no flat region. It is also stated in \cite{hobson-epjc-2022} that, since CGR is (by construction) conformally invariant, conformal transformations should not change the ``observable predictions''. This makes clear that the authors consider that only conformal invariant quantities, in particular the conformal invariant metric $\mathfrak{g}_{\mu\nu}\equiv(\phi/\sqrt{6}M_\text{pl})^2g_{\mu\nu}$, have observational meaning. However, since according to the passive approach to Weyl symmetry the Weyl-invariant metric coincides with the GR-dS metric: $\mathfrak{g}_{\mu\nu}=(\phi/\sqrt{6}M_\text{pl})^2g_{\mu\nu}=\hat g_{\mu\nu}$, only the GR-SdS geometry \eqref{schw-ds-gr} has an observational meaning. 

We must recall that this analysis is correct only if one assumes from the start that the passive Weyl transformations represent the only correct approach to conformal symmetry. As we have shown, if we follow this approach, the action and derived EOM of CGR theory, when written in terms of the physical quantities, coincide with the GR action and derived EOM, so Weyl symmetry is lost or is, at least, a fictitious symmetry. Recall that in this case the ``auxiliary'' fields that suffer the Weyl transformations: $g_{\mu\nu}$, $\phi$, $\chi_i$, and $m$ are removed from the covariant formulation \eqref{tot-lag-pwt}.



\section{Weyl-invariant cosmology and the many-worlds perspective}
\label{sect-cosmo}


The CGR theory \eqref{tot-lag} is a much more versatile alternative to solving current gravitational puzzles than GR alone, since the latter belongs to the conformal equivalence class $\mathfrak{C}_\text{conf}$ of CGR. Suppose that we have an arbitrary GR solution labeled ``$A$'' (from now on the upper case latin indexes $A$, $B$, $C$, etc., run over all possible solutions of GR-EOM; $A=1,2,3,...,N_s$, where $N_s$ is the number of different solutions), and that a corresponding global gravitational state labeled ``$k$'' is represented as (compare with \eqref{gauge-def});

\begin{align} \mathfrak{G}(A|k):\{\phi_k(x),g^{(A|k)}_{\mu\nu}(x),...\},\;\;A=1,2,...,N_s;\;\;k=1, 2,..., N,\label{k-gauge}\end{align} where $N\rightarrow\infty$, and the ellipsis in curly brackets stands for other fields coupled to gravity. Let us assume that the fields $\phi_k(x)$ and $g^{(A|k)}_{\mu\nu}(x)$ satisfy the CGR-EOM \eqref{einst-eom'} for any $k$ and any gauge choice $\phi_k=\phi_k(x)$. The different metric tensors $g^{(A|k)}_{\mu\nu}(x)$ are related to the GR solution labeled ``$A$'' through \eqref{invert-gauge-t};

\begin{align} g^{(A|k)}_{\mu\nu}(x)=\left(\frac{\sqrt{6}M_\text{pl}}{\phi_k(x)}\right)^2\hat g^{(A)}_{\mu\nu}(x),\nonumber\end{align} where $\hat g^{(1)}_{\mu\nu}(x)$, $\hat g^{(2)}_{\mu\nu}(x)$,...,$\hat g^{(N_s)}_{\mu\nu}(x)$, are different solutions of the GR-EOM. Hence, instead of a single GGS: $\mathfrak{G}_\text{gr}(A):\{\sqrt{6}M_\text{pl},\hat g^{(A)}_{\mu\nu}(x),...\}$, corresponding to each GR solution ``$A$'', in CGR there is an infinite set of global gravitational states: $\mathfrak{C}_\text{conf}(A)=\{\mathfrak{G}_\text{gr}(A),\mathfrak{G}(A|1),\mathfrak{G}(A|2),...,\mathfrak{G}(A|N-1)\}$ ($A=1,2,...,N_s$), including the GR solution itself, which are available to describe a given set of gravitational phenomena. Only experiments can pick out the GGS in $\mathfrak{C}_\text{conf}(A)$ that more closely describes given phenomena.

Let us illustrate the versatility of CGR in the cosmological framework. Here, we assume a Friedmann-Robertson-Walker (FRW) line-element with flat spatial sections:

\begin{align} ds^2=-dt^2+a^2(t)\delta_{ij}dx^idx^j,\label{line-e}\end{align} where $t$ is cosmic time, $a(t)$ is the scale factor, and the lower case latin indices run through space: $i,j=1,2,3$. We further assume that the background matter content is a perfect fluid with energy density $\rho$ and pressure $p$ that satisfy the following EOS: $p=(\gamma-1)\rho$, where $\gamma$ is the barotropic index of the fluid. The independent equations are the $(0,0)$-component of the Einstein-type EOM \eqref{einst-eom'} and the nonhomogeneous continuity equation \eqref{cont-eq};

\begin{align} \left(\frac{a'}{a}+\frac{\phi'}{\phi}\right)^2&=\frac{2a^2}{\phi^2}\,\rho,\label{fried-eom}\\
\rho'+3\gamma\frac{a'}{a}\,\rho&=(4-3\gamma)\frac{\phi'}{\phi}\,\rho,\label{nhom-cont-eom}\end{align} respectively, where the tilde accounts for the derivative with respect to the conformal time $\tau=\int dt/a(t)$.\footnote{We use the conformal time instead of the cosmic time $t$ because the former is a Weyl-invariant quantity, while $dt\rightarrow d\hat t=\Omega\,dt$ under WTs.} Straightforward integration of \eqref{nhom-cont-eom} yields $\rho=(C^2/2)\,\phi^4(a\phi)^{-3\gamma}$ where $C^2$ is an integration constant. Taking into account this equation, if we introduce the Weyl-invariant variable $v\equiv a\phi$, the Friedmann EOM \eqref{fried-eom} can be written as $v'=C\,v^{\frac{4-3\gamma}{2}}$. This is a Weyl-invariant equation whose integration yields $v(\tau)=\bar C\,(\tau-\tau_0)^\frac{2}{3\gamma-2}$, where $\tau_0$ is another integration constant and $\bar C\equiv[(3\gamma-2)\,C/2]^{2/(3\gamma-2)}$. We have

\begin{align} a(\tau)=\bar C\,(\tau-\tau_0)^\frac{2}{3\gamma-2}\phi^{-1}(\tau).\label{sol}\end{align} Different choices of the gauge field: $\phi(\tau)=\phi_1(\tau)$, $\phi(\tau)=\phi_2(\tau)$,...,$\phi(\tau)=\phi_N(\tau)$ ($N\rightarrow\infty$), lead to different behaviors of the scale factor: $a_1(\tau)$, $a_2(\tau)$,..., $a_N(\tau)$. In the GR gauge, we have: $\phi(\tau)=\phi_\text{gr}=\sqrt{6}M_\text{pl}$, so 

\begin{align} a_\text{gr}(\tau)=\frac{\bar C}{\sqrt{6}M_\text{pl}}\,(\tau-\tau_0)^\frac{2}{3\gamma-2}.\nonumber\end{align} 

In CGR theory, instead of a single GGS for each value of the barotropic index $\gamma$, as in GR, we have an infinite class of global gravitational states:

\begin{align} \mathfrak{C}_\text{conf}=\left\{\left(\sqrt{6}M_\text{pl},a_\text{gr}(\tau)\right),\left(\phi_1(\tau),a_1(\tau)\right),...,(\phi_{N-1}(\tau),a_{N-1}(\tau))\right\}.\nonumber\end{align} These are related by the following Weyl transformation:

\begin{align} a_k(\tau)=\left[\frac{\sqrt{6}M_\text{pl}}{\phi_k(\tau)}\right]a_\text{gr}(\tau),\;\;k=0,1,2,...,N-1,\nonumber\end{align} where the label $k=0$ is used exclusively to denote the GR gauge. That is, $\phi_0=\sqrt{6}M_\text{pl}$. 

The large (in principle infinite) variety of cosmic behaviors in $\mathfrak{S}_\text{cgr}$ is what makes the CGR theory and, therefore, Weyl symmetry attractive within the cosmological context and, at the same time, yields the most unappealing feature of this theory when it is analyzed under the active approach: the loss of predictability. The Weyl-invariant CGR theory is not capable of defining a unique GGS, but a whole class of them, so that the definition of the state that correctly describes a given gravitational phenomenon is left to experiment. Nevertheless, this theory can be assigned some predictive power. For example, the fifth force $f^\mu=h^{\mu\lambda}\der_\lambda\phi/\phi$ in \eqref{timelike-eom}, which is associated with point-dependent masses so it can pass unnoticed in local experiments, points to a modification of the redshift measurements that could explain, in principle, the present accelerated pace of the cosmic expansion and so-called dark matter, because it does not interact with radiation ($T^\text{rad}=0$ $\Rightarrow$ $f_\mu=0$). In the last instance, the active approach to Weyl symmetry within CGR theory may be viewed as a classical version of the many-worlds interpretation of gravitational interactions of matter \cite{everett_rmp_1957, wheeler, dewitt_phys_rev_1967, dewitt, barvinsky_cqg_1990, omnes_rmp_1992, tegmark_1998, garriga_prd_2001, zurek_rmp_2003, tegmark_nature_2007, quiros_prd_2023}.

There is another issue that warrants discussion. To introduce the problem, we shall follow the line of reasoning of \cite{faraoni_prd_2007}. Consider the line-element in the GR gauge and in any other $k$-gauge: $ds^2_\text{gr}=a^2_\text{gr}(\tau)(-d\tau^2+\delta_{ij}dx^idx^j)$ and $ds^2_k=a^2_k(\tau)(-d\tau^2+\delta_{ij}dx^idx^j)$, respectively. According to the bibliographic reference mentioned above, the physically meaningful quantity to analyze, for instance, the singularity issue, is not the scale factor itself but a typical physical (proper) length $a(\tau)\sqrt{\delta_{ij}dx^idx^j}$. The corresponding measured quantity would be the ratio $a(\tau)\sqrt{\delta_{ij}dx^idx^j}/u(\tau)$, where $u$ is the unit of length. Since $u$ transforms in the same way as the proper length, then:

\begin{align} \frac{a_k(\tau)\sqrt{\delta_{ij}dx^idx^j}}{u_k(\tau)}=\frac{a_\text{gr}(\tau)\sqrt{\delta_{ij}dx^idx^j}}{u_\text{gr}}.\nonumber\end{align} This means that there is no measurable effect that the gauge choice could have. However, there are some flaws in this line of reasoning. For example, since here we follow the active approach to Weyl transformations (the only possible approach that leads to Weyl symmetry having phenomenological consequences \cite{quiros_prd_2025}), the GR metric $g^\text{gr}_{\mu\nu}(\tau)$ and any of its conformal metrics $g^{(k)}_{\mu\nu}(\tau)$ have independent physical meaning. In particular, the curvature tensors and any of their contractions are made up of the derivatives of these metric tensors. In the present cosmological example, the curvature scalar in the $k$-gauges reads: $R_k=6(\dot H_k+2H^2_k)$, while in the GR gauge: $R_\text{gr}=6(\dot H_\text{gr}+2H^2_\text{gr})$, where the Hubble parameter is defined as $H_k=\dot a_k/a_k$, $H_\text{gr}=\dot a_\text{gr}/a_\text{gr}$, etc. Here, the dot accounts for the derivative with respect to cosmic time $t$, so $dt=ad\tau$. Under WTs, the Hubble parameter transforms as $H\rightarrow H_\text{gr}=\Omega^{-1}(H+\dot\phi/\phi)$. The transformation of $\dot H$ is even more complex. So is the transformation of the curvature scalar. Hence, the arguments of \cite{faraoni_prd_2007} do not apply when the derivatives of the scale factor are implied. 

Additionally, many measurements in cosmology rest on the emission and detection of photons. These follow null-geodesics of Riemann geometry, which are themselves form-invariant under WTs. Any measuring clock or stick, which is made of a combination of emitted and/or absorbed photons, is not transformed by the Weyl transformations and so is the related measuring unit. Hence, the above argument \cite{faraoni_prd_2007} is not universal and can be circumvented.



\section{Observational consequences of CGR: The redshift measurements}
\label{sect-observ}


Even if measuring units, for example, the mass unit, are point-dependent, there are ways in which this effect can be experimentally revealed. For example, the point-dependent property of the masses of timelike fields, and so the fifth force appearing in \eqref{cont-eq}, cannot be detected in local experiments because any measuring stick is made up of fields, so it suffers the same universal point-dependence as the mass being measured, and the effect cancels out. However, in experiments that involve the detection of photons emitted at a distant point in spacetime, namely redshift measurements, at least in principle, the effect can be measured.

Let us explain the peculiar contribution to the redshift of frequencies carried by the Weyl symmetry. In CGR theory, two phenomena contribute to the redshift of frequencies: 1) the gravitational redshift (also curvature redshift) and 2) the redshift due to variation of masses in spacetime (mass redshift). The former effect is due to the propagation of photons in a curved background, while the latter is due to the point-dependent property of masses: $m=m(x)=\kappa\,\phi(x)$. This mass dependence affects the redshift because the frequency of a photon emitted/absorbed by some atom (atomic number $Z$) is given by the Rydberg formula:

\begin{align} \gamma_{e/a}=Z^2\frac{m_ec^2\alpha^2}{2h}\left(\frac{1}{n^2_l}-\frac{1}{n^2_u}\right),\label{rydberg-f}\end{align} where $m_e$ is the mass of the electron, $c$ is the speed of light, $h$ is the Planck constant (here, temporarily we assume that $c\neq h\neq 1$), $\alpha$ is the fine-structure constant ($\alpha\approx 1/137$), while $n_l$ and $n_u$ are the principal quantum numbers of the lower and upper energy levels, respectively. Consider two identical atoms, one of which emits a photon at some time $t$ in the past, and the other atom absorbs the photon at present time $t_0$. The relative shift of the frequency of the emitted/absorbed photons is given by

\begin{align} z_m=\frac{\gamma_{e/a}(t)-\gamma_{e/a}(t_0)}{\gamma_{e/a}(t_0)}=\frac{m_e(t)-m_e(t_0)}{m_e(t_0)}=\frac{\phi(t)}{\phi(t_0)}-1.\label{z-mass}\end{align} Note that this emission/absorption frequency does not depend on the specific atom that emits/absorbs the photon, nor on the specific atomic transition $n_u\rightleftharpoons n_l$. Note also that in the GR gauge, since $\phi(t)=\phi(t_0)=\sqrt{6}M_\text{pl}$, this type of redshift does not arise. Meanwhile, the curvature redshift is due to the dynamics of the photon frequency during its propagation in a curved background. The frequency of a photon at any time $t$ while it propagates in an expanding universe can be found as a solution of the null-geodesics \eqref{0-geod}: $\gamma(t)=\gamma_0/a(t)$, where $\gamma_0$ is some constant frequency. The curvature redshift resulting from the propagation of a photon in an expanding (homogeneous and isotropic) FRW background between times $t$ and $t_0$, and the overall redshift are given by:

\begin{align} z_c=\frac{\gamma(t)-\gamma(t_0)}{\gamma(t_0)}=\frac{a(t_0)}{a(t)}-1=\frac{\phi(t)}{\phi(t_0)}\frac{\hat a(t_0)}{\hat a(t)}-1,\label{z-curv}\end{align} and 

\begin{align} z_\text{tot}=z_c+z_m=\frac{a(t_0)}{a(t)}+\frac{\phi(t)}{\phi(t_0)}-2=\frac{\phi(t)}{\phi(t_0)}\frac{\hat a(t_0)}{\hat a(t)}+\frac{\phi(t)}{\phi(t_0)}-2,\label{z-tot}\end{align} respectively, where we took into account the relationship, $a(t)=(\sqrt{6}M_\text{pl}/\phi(t))\,\hat a(t)$, with $\hat a$ being the scale factor in the GR gauge. If we realize that the GR redshift $z={\hat a(t_0)/\hat a(t)}-1$, then

\begin{align} z_c+1=\frac{\phi(t)}{\phi(t_0)}(z+1),\;z_\text{tot}+2=\frac{\phi(t)}{\phi(t_0)}(z+2).\nonumber\end{align} In a cosmological context, for purpose of comparizon with observations, it is better to replace the cosmic time $t$ by the redshift $z$. This is possible thanks to the relationship $\hat a(t)=\hat a(t_0)/(z+1)$, from where we can have $t=f(z)$. Hence, $\phi(t)=\phi(f(z))=\vphi(z)$, where $\vphi=\phi\circ f$. We have,

\begin{align} z_c+1=\frac{\vphi(z)}{\vphi(0)}(z+1),\;z_\text{tot}+2=\frac{\vphi(z)}{\vphi(0)}(z+2),\label{cosmo-z}\end{align} where we assumed that $t_0$ is the present time, so that $t_0=f(z)|_0=f(0)$ $\Rightarrow \phi(t_0)=\vphi(0)$.

Let us, for simplicity and definiteness, choose the linear $z$ gauge

\begin{align} \frac{\vphi(z)}{\vphi(0)}=1+\frac{\vphi'(0)}{\vphi(0)}z=1+\mu z,\label{z-approx}\end{align} where a prime means derivative with respect to $z$ and $\mu$ is a dimensionless constant. Hence, in this ``linear gauge'' 

\begin{align} 1+z_c=(1+\mu z)(1+z),\;z_\text{tot}=(1+2\mu+\mu z)z.\label{z-approx'}\end{align} 

The expressions in \eqref{z-approx'} show that in this gauge of CGR cosmology, both the curvature redshift $z_c$ and the measured redshift $z_\text{tot}$ are greater than the redshift according to the GR gauge. This result can have obvious implications for the explanation of the late-time dynamics of our Universe. Since the fifth force that arises in \eqref{cont-eq} and in \eqref{timelike-eom} is due to the point-dependent property of the mass: $m(x)=\kappa\,\phi(x)$, accurate measurements of the relationship on the right in \eqref{cosmo-z} would entail indirect measurements of the extra force $f^\alpha=h^{\alpha\lambda}\der_\lambda\phi/\phi$ because, if instead of $z_\text{tot}+2=[\vphi(z)/\vphi(0)](z+2)$, a relationship $z_\text{tot}+1=[\vphi(z)/\vphi(0)](z+1)$ is verified, this would mean that there is no redshift associated with the mass variation and, so, $m$ must be a constant; that is $m\to\Omega^{-1}m$ is not correct.


\subsection{SNIa redshift measurements: Weyl symmetry vs dark energy}


In view of the redshift measurements of supernovae type Ia \cite{riess-1998, perlmutter-1999, desi}, the distances of the high-redshift SNIa are farther than expected in a low mass density universe described by GR theory without a cosmological constant. This finding triggered the need for DE, with the $\Lambda$-cold dark matter ($\Lambda$CDM) model serving as the reference. Let us discuss how the SNIa redshift measurements reported in the above bibliographic references can be explained as a natural consequence of Weyl symmetry.

Measuring distances to distant stars is a challenging task. Several quantities, such as the apparent magnitude $m$, the absolute magnitude $M$ (logarithmic measures of flux and luminosity, respectively), and the luminosity distance $d_L$ are involved \cite{ccp-padman, dodelson-book}. The latter is related to the energy flux $\Phi$ measured by the observer at $z=0$, which comes from a distant source with actual luminosity $L$; $d_L=\sqrt{L/4\pi\Phi}$. It is, in general, a model-dependent quantity. It can be related to theoretical quantities as it follows \cite{dodelson-book}:

\begin{align} d_L=\frac{d_C}{a}=(1+z)d_C,\label{dl}\end{align} where the comoving distance $d_C$ is given by

\begin{align} d_C=\int_t^0\frac{dt'}{a(t')}=\int_0^z\frac{dz'}{H(z')}.\label{co-mov-d}\end{align} The expression for the Hubble parameter as a function of the redshift parameter $z$ can be found from the Friedmann equation of the model. For example, for the $\Lambda$CDM model we have:

\begin{align} H(z)=H_0\sqrt{\Omega_{m,0}(z+1)^3+\Omega_{\Lambda,0}},\label{lcdm}\end{align} where the dimensionless energy density of matter and of the cosmological constant are given by: $\Omega_{m,0}\equiv\rho_{m,0}/3H^2_0$, and $\Omega_{\Lambda,0}\equiv\Lambda/3H^2_0$, respectively. In the above equations, $H_0$ and $\rho_{m,0}$ are the present values of the Hubble parameter and of the matter density. 

For simplicity, we shall investigate the modification to the redshift in the linear gauge \eqref{z-approx} where the relationship \eqref{z-approx'} between the overall (measured) redshift in this gauge and the GR redshift $z$ takes place. Note that in the expression for the comoving distance $d_C$, the expression under the integral: $dt/a(t)$ is a Weyl form-invariant quantity, so it has the same functional dependence in the linear gauge and in the GR gauge ($\Lambda=0$, $\Omega_{m,0}=1$). That is, $\int_t^0 dt'/a(t')=\int_0^zdz'/(z'+1)^{3/2}$. However, a subtlety must be considered: while the integration limit in GR gauge is $z$, in the ``linear gauge'' it must be $z_\text{tot}$, which is the measured redshift in that gauge. This means that in GR gauge we have:

\begin{align} H_0d_C=\int_0^z\frac{dz'}{(z'+1)^{3/2}}=2\left(\frac{\sqrt{z+1}-1}{\sqrt{z+1}}\right),\label{dc}\end{align} while in the linear gauge \eqref{z-approx'} we get that

\begin{align} H_0d^\text{lin}_C=\int_0^{z_\text{tot}}\frac{dz'}{(z'+1)^{3/2}}=2\left(\frac{\sqrt{z_\text{tot}+1}-1}{\sqrt{z_\text{tot}+1}}\right)=2\left(1-\frac{1}{\sqrt{1+(1+2\mu+\mu z)z}}\right),\label{dc'}\end{align} where $z_\text{tot}$ is given by the expression on the right in \eqref{z-approx'}. 

Another subtlety to be considered is that, according to \eqref{z-curv} and \eqref{dl}, in an arbitrary gauge $d_L=d_C/a=d_C(1+z_c)$. That is, only the curvature redshift matters, since it is the one related to the scale factor \eqref{z-curv}. Hence, in the linear gauge \eqref{z-approx}, the expression for the luminosity distance is given by:

\begin{align} H_0d^\text{lin}_L=(1+z_c)H_0d^\text{lin}_C=(1+\mu z)(1+z)H_0d^\text{lin}_C,\nonumber\end{align} where $H_0d^\text{lin}_C$ is given by \eqref{dc'} and we took into account the expression for $1+z_c$ on the left in \eqref{z-approx'}.

According to Eq. (31) of reference \cite{ccp-padman}, the redshift-dependent apparent magnitude $m$ of distant SNIa is given by\footnote{Since we consider the dimensionless combination $H_0d_L$ rather than $d_L$, this means that in the equation to determine the apparent magnitude a term $5\log_{10}h$ is to be taken away, where $H_0=100h\,\text{km}\,\text{s}^{-1}\,\text{Mpc}^{-1}$.}

\bea m=M-5\log_{10}h+42.38+5\log_{10}\left(H_0d_L\right).\nonumber\eea Here we take $M=-19.09$ and $h=0.72$ so that,

\bea m=5\log_{10}(H_0d_L)+24,\label{app-m}\eea which is the master equation to determine the theoretical values of the apparent magnitude.


\begin{figure*}[t!]
\includegraphics[width=8cm]{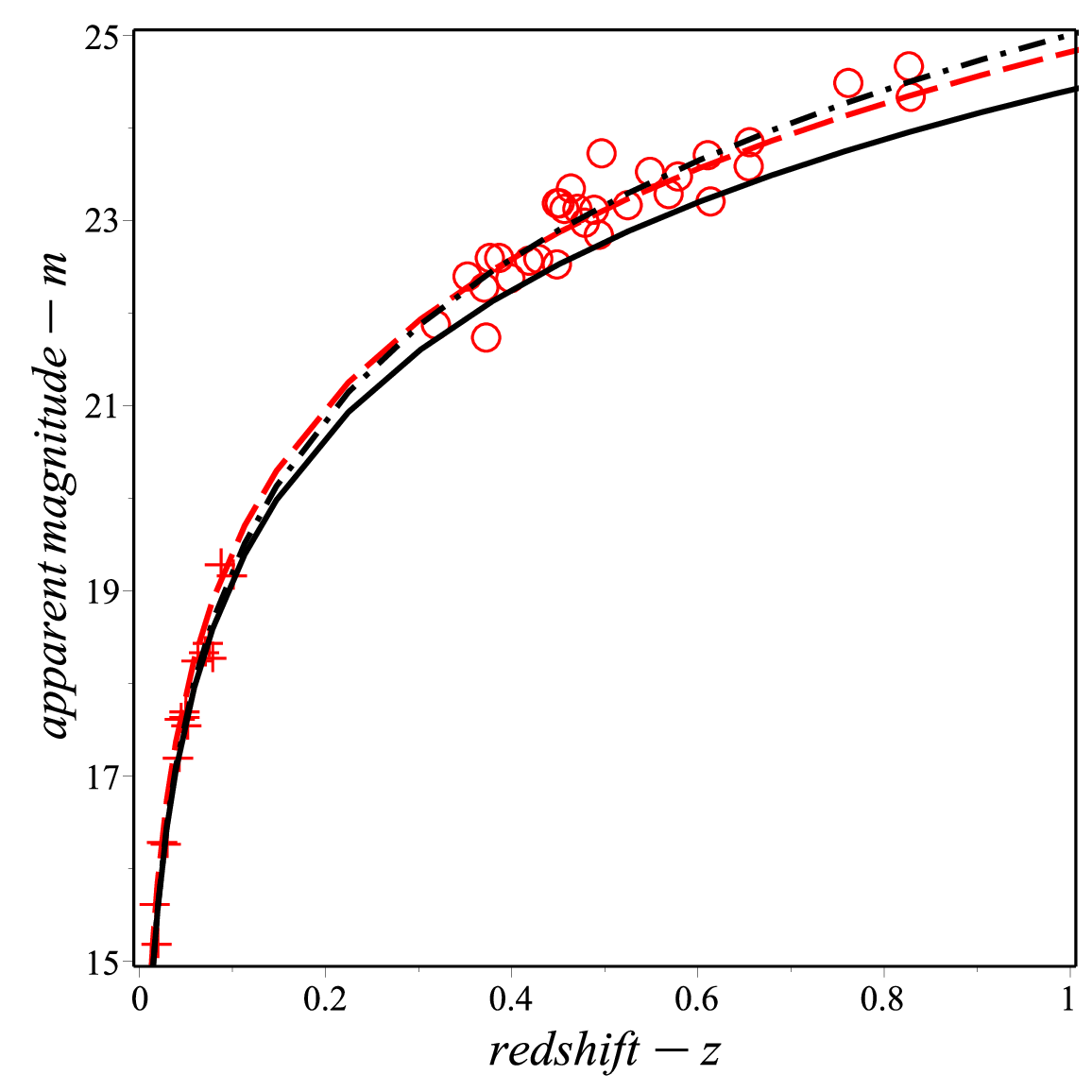}
\includegraphics[width=8cm]{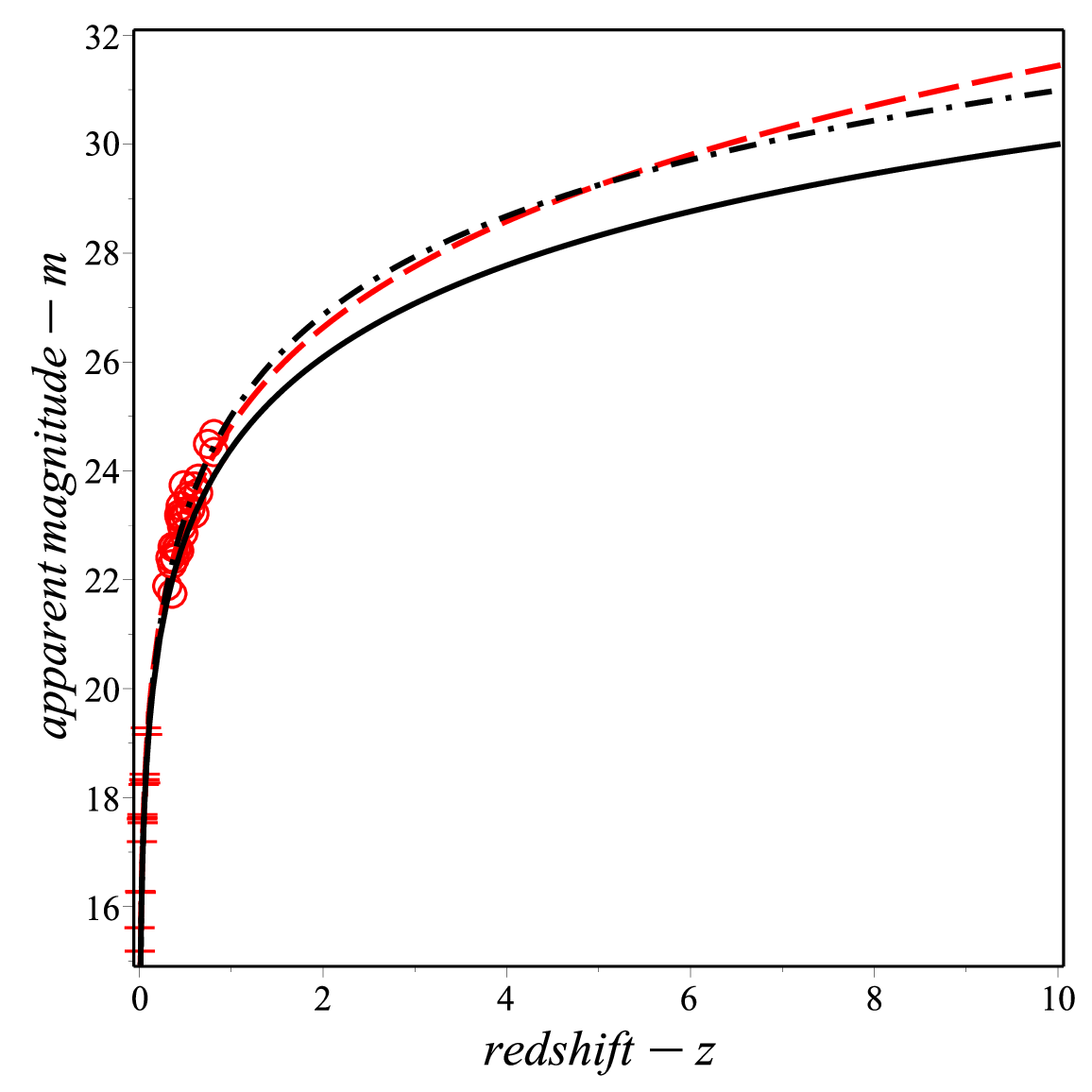}
\vspace{1.3cm}\caption{Plots of the apparent magnitude $m$ \eqref{app-m} vs redshift $z$, computed for three different cases: 1) GR model ($\Lambda=0$) \eqref{case-gr}, 2) the reference $\Lambda$CDM model \eqref{case-lcdm} (the values $\Omega_{m,0}=0.3$ and $\Omega_{\Lambda,0}=0.7$ have been chosen), and 3) the Weyl-invariant CGR model \eqref{case-cgr}. In the plots, the solid curve corresponds to the GR model (vanishing cosmological constant), while the dash-dot curve corresponds to the reference model ($\Lambda$CDM) and the red dashed curve corresponds to the Weyl-invariant CGR model \eqref{case-cgr} with $\mu=0.075$. The red crosses and the circles represent observational points corresponding to small-redshift and high-redshift data, taken from TAB. 2 and TAB. 1 of Ref. \cite{perlmutter-1999}, respectively. We have not included all data points, but just a representative set of them. The error bars have not been included since the plots are for illustrative purposes. In the figure on the right, the redshift interval is widened: $z\in[0,10]$, to show how the CGR model appreciably departs from the reference $\Lambda$CDM model at redshifts larger than $z\approx 6$.}\label{fig1}\end{figure*}


Now we are in a position to explain how the accelerated expansion is naturally avoided in the present conformal invariant theory, due to the difference of the measured redshift in the Weyl invariant CGR and in GR theory. As an illustration, we shall consider one of the first data sets on high redshift supernovae measurements \cite{perlmutter-1999}, which provided early evidence for accelerated expansion of the universe. We shall consider three cases: 1) the GR model, where

\begin{align} H_0d_L=(1+z)H_0d_C,\label{case-gr}\end{align} and $H_0d_C$ is given by \eqref{dc}, 2) the GR $+\Lambda$ ($\Lambda$CDM) model, where

\begin{align} H_0d^\Lambda_L=(1+z)H_0d^\Lambda_C,\;H_0d^\Lambda_C=\int_0^z\frac{dz'}{\sqrt{\Omega_{m,0}(z'+1)^3+\Omega_{\Lambda,0}}},\label{case-lcdm}\end{align} and 3) the Weyl-invariant CGR model:

\begin{align} H_0d^\text{lin}_L=(1+\mu z)(1+z)H_0d^\text{lin}_C,\label{case-cgr}\end{align} where $H_0d^\text{lin}_C$ is given by \eqref{dc'}. We compute the theoretical values of the apparent magnitude $m$ \eqref{app-m} for the three cases and compare the results. These are checked against redshift data. 

The results are shown in FIG. \ref{fig1}. It is seen that, while the GR gauge does not explain the SNIa luminosity distance dataset, the simple linear gauge \eqref{z-approx} of Weyl-invariant CGR yields as good a fitting of the SNIa observational data as the $\Lambda$CDM reference model. For $z\gtrsim 6$, both models start departing from each other.

The above analysis can be summarized as follows. Although there is an infinity of other gauges of CGR theory, for illustrative purposes, here we considered only two different gauges: $\mathfrak{G}_\text{gr}$ and $\mathfrak{G}_\text{lin}$. These belong in the same CEC $\mathfrak{C}_\text{conf}$ \eqref{c-class}. In addition, for comparison purposes exclusively, we also included in the analysis the reference $\Lambda$CDM model, which shows a good fit of the SNIa data on luminosity distance measurements. This model is not in the same equivalence class with $\mathfrak{G}_\text{gr}$ and $\mathfrak{G}_\text{lin}$. It is included to show that the addition of a cosmological constant term in the GR theory can account for the present accelerated expansion inferred from SNIa observational data. The GR gauge, in which the luminosity distance is given by \eqref{case-gr} and represented by the solid black curve in FIG. \ref{fig1} cannot explain the SNIa observational data (the red crosses and circles in the figure). In contrast, the linear gauge with the luminosity distance $d^\text{lin}_L$ given by \eqref{case-cgr} and represented by the dashed red curve in FIG. \ref{fig1}, explains the observational data on SNIa luminosity distance, showing as good a fit as the reference $\Lambda$CDM model. Hence, the Weyl-invariant CGR theory explains the luminosity distance observational data as a consequence of the measured redshift in a gauge different from the GR gauge. In this regard, we may consider the observational data as a means to pick out a gauge among the infinity of possible gauges in $\mathfrak{C}_\text{conf}$, that represent a potential description of the gravitational phenomena in this theory. In the present case, among the two gauges considered: the GR gauge and the simple linear gauge, the observational data shows better fit to the latter gauge.



\section{Quantum-mechanical removal of spacetime singularities}
\label{sect-sing}


Let us discuss the possibility that quantum-mechanical considerations within the CGR framework may resolve the spacetime-singularity issue; that is, that quantum theory may remove the spacetime singularities that plague classical gravitational theory \cite{penrose-1, penrose-2, penrose-3, hawking-1, berger, bosma}. According to the previous discussion in sections \ref{sect-many-w} and \ref{sect-cosmo}, a global gravitational state $\mathfrak{G}(A|k)=\{\phi_k(x),g^{(A|k)}_{\mu\nu},...\}$ can be thought as a fourth-dimensional sheet where the whole history of all fields is drawn: the metric tensor $g^{(A|k)}_{\mu\nu}=g^{(A|k)}_{\mu\nu}(t,\vec{x})$, the scalar field $\phi_k=\phi_k(t,\vec{x})$, and each matter field coupled to gravity $\chi^{(A)}_i=\chi^{(A)}_i(t,\vec{x})$ (for simplicity we assume that the matter fields are fixed, that is, these are the same in any gauge). Each sheet is labeled by the pair $(A|k)$, i.e., by the index corresponding to the given GR solution $A=1,2,...,N_s$, and by the index corresponding to the gauge choice $k=1,2,...,N$ ($N\rightarrow\infty$). Let us call it a ``gauge sheet''.\footnote{The gauge sheet is just a vague illustration of the actual meaning of a gauge or what we call here a global gravitational state. A clearer illustration must be consistent with the well-posed initial-value formulation of CGR theory. We assume that such a gauge sheet, for example, the one associated to an arbitrary solution ``$A$'' and a gauge choice $\phi_k=\phi_k(x)$, is a globally hyperbolic spacetime $({\cal M},g^{(A|k)}_{\mu\nu})_{\phi_k}$ admitting foliation by Cauchy hypersurfaces $\sum_t$, parametrized by a global time function $t$. We may view this globally hyperbolic spacetime as representing the time development of a Riemannian metric on a fixed three-dimensional manifold \cite{wald-book}.} A given classical gravitational phenomenon is well-described by a single gauge sheet within the infinite collection of $N\to\infty$ gauge sheets, but quantum gravity must take into account all of them, since all of the possible gauges of a given solution contribute to the probability amplitude;\footnote{Although the probability amplitude of some solution, in particular \eqref{p-amp}, is expressible as \cite{hawking-2, hawking-3, hawking-4}: $$\vphi=\int {\cal D}\phi\,{\cal D}g\,\exp{iS_\text{cgr}[\phi,g]},$$ where ${\cal D}\phi$ and ${\cal D}g$, are some measures in the space of all scalar fields $\phi=\phi(x)$ and metrics $g_{\mu\nu}=g_{\mu\nu}(x)$, respectively, and $S_\text{cgr}[\phi,g]$ is the classical action of vacuum CGR theory \eqref{caction-cgr}, for our qualitative discussion it will be more adequate to work with the sum \eqref{p-amp}.}

\begin{align} \vphi(A)=\lim_{N\to\infty}\sum_{k=1}^N\vphi(A|k),\;\;\vphi(A|k)=\frac{\vphi_0}{N}\exp iS_\text{cgr}[\phi_k,g_{(A|k)}],\label{p-amp}\end{align} where $\vphi(A)$ is the probability amplitude of the solution ``$A$'', while $\vphi(A|k)$ is the probability amplitude of the gauge ``$k$'' of solution ``$A$'', $\vphi_0/N$ is some normalization constant, and the sum is over all gauges in the conformal class of the GR solution ``$A$'': $\mathfrak{C}_\text{conf}(A)=\{\mathfrak{G}_\text{gr}(A),\mathfrak{G}(A|1),\mathfrak{G}(A|2),...,\mathfrak{G}(A|N-1)\}$. For simplicity of the analysis, we consider the vacuum CGR theory. Hence,

\begin{align} S_\text{cgr}[\phi,g]=\frac{1}{12}\int_{{\cal V}_4}d^4x\sqrt{-g}\left[\phi^2R+6(\der\phi)^2\right]+\frac{1}{6}\int_{\der{\cal V}_4}d^3x\sqrt{-h}\phi^2K,\label{caction-cgr}\end{align} where ${\cal V}_4$ is some integration volume, $h_{\mu\nu}=g_{\mu\nu}\pm n_\mu n_\nu$ is the metric induced on the boundary of the integration volume  $\der{\cal V}_4$, which is orthogonal to the unit vector field $n_\mu$, and $K=h^{\mu\nu}K_{\mu\nu}$ ($K_{\mu\nu}=h^\lambda_{\;\mu}h^\sigma_{\;\nu}\nabla_\lambda n_\sigma$ is the extrinsic curvature tensor of the boundary). The boundary term is introduced to compensate for the second derivatives of the gravitational action in the variational procedure \cite{hawking-2}. This term does not modify the conformal form-invariance of the action \eqref{caction-cgr} (see, for example, appendix A of \cite{wands-rev}). 

Due to form-invariance of the classical action $S_\text{cgr}[\phi,g]$ under Weyl rescalings \eqref{aweyl-t}, each probability amplitude $\vphi(A|k)$ in \eqref{p-amp} contributes not only a same amplitude $\vphi_0/N$, but also a same phase $S_\text{cgr}$, so that $\vphi(A|1)=\vphi(A|2)=\vphi(A|*)=...=\vphi(A|N-1)$, where $\mathfrak{G}(A|*)$ can be any of the gauges in the CEC $\mathfrak{C}_\text{conf}(A)$. Therefore,

\begin{align} \vphi(A)=\vphi_0\exp iS_\text{cgr}[\phi_*,g_{(A|*)}].\label{p-amp'}\end{align}  

To explain how the spacetime singularity is removed in quantum-mechanical computations, let us consider a specific solution. Here, for illustration, we consider the GR vacuum Schwarzschild solution \eqref{schw-gr}:

\begin{align} d\hat s^2=\hat g^{(\cancel{0})}_{\mu\nu}dx^\mu dx^\nu=-\left(1-\frac{2m}{r}\right)dt^2+\frac{dr^2}{\left(1-\frac{2m}{r}\right)}+r^2d\Omega^2.\nonumber\end{align} Any conformal Schwarzschild metric \eqref{conf-schw-gr}

\begin{align} ds^2=g^{(\cancel{0}|*)}_{\mu\nu}dx^\mu dx^\nu=\left[\frac{\sqrt{6}M_\text{pl}}{\phi_*(r)}\right]^2\left[-\left(1-\frac{2m}{r}\right)dt^2+\frac{dr^2}{\left(1-\frac{2m}{r}\right)}+r^2d\Omega^2\right],\label{conf-schw-met}\end{align} is also a solution of the vacuum CGR-EOM \eqref{vac-eom}. For example, in \cite{bambi-2017-a}, the authors showed that the gauge choice $\phi_*(r)=\sqrt{6}M_\text{pl}/\sqrt{1+(L/r)^4},$ where $L$ is some length scale, leads to the Schwarzschild conformal metric being complete, so the resulting geometry represents a nonsingular black hole. According to \eqref{p-amp'}, the probability amplitude of the Schwarzschild solution may be given by: $\vphi(\cancel{0})=\vphi_0\exp iS_\text{cgr}[\phi_*,g_{(\cancel{0}|*)}],$ where $\phi_*$ corresponds to the gauge choice made in \cite{bambi-2017-a} (see above) and $g^{(\cancel{0}|*)}_{\mu\nu}$ is the conformal Schwarzschild metric \eqref{conf-schw-met}, which is complete, so there will be no singularity issue. Due to conformal form-invariance of the classical CGR action, leading to the same phase contribution to the probability amplitude from any gauge, we can choose any gauge, say $\mathfrak{G}(A|*)$ $\Rightarrow\,S_\text{cgr}[\phi_*,g_{(\cancel{0}|*)}],$ where the singularity is removed. This is the way in which the spacetime singularities are avoided in quantum-mechanical computations.

A final comment on this issue: Within the path integral formulation of gravity, not only those fields that are solutions of the gravitational EOM, like, for example, $\phi_*$ and $g^{(\cancel{0}|*)}_{\mu\nu}$, matter. Any field configurations that do not satisfy the gravitational EOM are to be considered as well. Notice, however, that the classical action $S_\text{cgr}[\phi,g]$ itself is conformal form-invariant regardless of whether the fields satisfy equations of motion or not. That is, any probability amplitudes contribute the same amplitude and phase. Hence, we are free to choose one of the singularity-free configurations and substitute it in \eqref{p-amp'}, even if the fields do not satisfy any equation of motion.



\section{Concluding remarks}
\label{sect-discu}


In this document, we have demonstrated that the well-known result in the HEP community that local scale invariance in flat and, more generally, in curved spacetime requires the vanishing trace of the matter stress-energy tensor: $T^{(m)}=0$, must be contrasted with the absence of this requirement when Weyl symmetry takes place, instead. In other words, the requirement $T^{(m)}=0$ holds whenever the masses of timelike fields are constant parameters that are not affected by Weyl rescalings. In the case of perfect fluids, this result means that, under WTs, the energy density of the fluid (the same as that of its pressure) transforms as $\rho\rightarrow\Omega^{-3}\rho$. Therefore, the Lagrangian density of matter fields and/or of perfect fluids is not form-invariant under the conformal transformations. In contrast, if the masses of timelike fields are point-dependent quantities transforming as $m\rightarrow\Omega^{-1}m$, under the WTs, and the energy density of perfect fluids transforms as $\rho\rightarrow\Omega^{-4}\rho$, the Lagrangian density of matter is indeed form-invariant under Weyl transformations. In this case, Weyl symmetry does not require the vanishing trace of the matter SET. Actually, when the matter field has a nonvanishing mass, which is transformed under \eqref{gauge-t} as $m\rightarrow\Omega^{-1}m$, the matter Lagrangian density is form-invariant under WTs \eqref{aweyl-t}, so that the Ward identity \eqref{mat-ward-id},

\begin{align} g^{\mu\nu}\frac{\delta{\cal L}_m}{\delta g^{\mu\nu}}=-\frac{\phi}{2}\frac{\delta{\cal L}_m}{\delta\phi}\;\Rightarrow\;\frac{\delta{\cal L}_m}{\delta\phi}=\frac{\sqrt{-g}}{\phi}\,T^{(m)},\nonumber\end{align} takes place. This variational derivative provides the term with the nonvanishing trace $T^{(m)}$ in the $\phi$-EOM \eqref{correct-kg-eom}: 

\begin{align} \frac{\delta{\cal L}_\text{grav}}{\delta\phi}+\frac{\delta{\cal L}_m}{\delta\phi}=-\sqrt{-g}\left[\nabla^2\phi-\frac{\phi}{6}R-\frac{1}{\phi}T^{(m)}\right]=0.\nonumber\end{align} Therefore, the $\phi$-EOM coincides with the trace of the Einstein-type EOM. That is, KG-type EOM \eqref{correct-kg-eom} is not an independent equation, so it does not contribute to the dynamics of the CGR theory. Our result that under the WTs \eqref{aweyl-t}, vanishing of the trace of the stress-energy tensor of matter is not required for Weyl invariance to be a symmetry of theory \eqref{tot-lag}, may be of importance for a revised perspective of the Weyl anomaly \cite{capper-duff-1974, duff-1977, rusos-1984, affleck-1986, duff-1994, odintsov-2000, cai-2010}. Unfortunately, the investigation of this issue clearly goes beyond the scope of the present document.


The above result indicates that Weyl invariance can be a genuine symmetry of the classical laws of gravity, implying that this symmetry must play a significant role in gravitational phenomenology. The fact that a fifth force inevitably arises if the gravitational laws are Weyl-invariant is a distinctive feature of CGR. As shown, the fifth force acts selectively only on timelike fields in the same manner as dark matter and dark energy do. This hints at a plausible explanation of the dark cosmological sector as a consequence of Weyl symmetry. We have been able to explain, in particular, the observational data on luminosity distance of SNIa \cite{riess-1998, perlmutter-1999, desi}, without resorting to dark energy. However, the present analysis has been mostly illustrative, and it did not involve other data sets than one of the first high-redshift measurements reported in \cite{perlmutter-1999}. Other evidence for accelerating expansion within the framework of GR-based $\Lambda$CDM model, such as cosmic microwave background (CMB) temperature anisotropies, baryon acoustic oscillations (BAO), etc., should be carefully analyzed within the framework of our present theory before we may come to a definitive (solid) conclusion on the possible abandonment of the dark energy idea. The latter studies require the development of perturbative analysis, which is beyond the aims of the present document.


The results discussed in this document are valid only for the Weyl-invariant CGR theory \eqref{tot-lag}. Consider, for example, conformal gravity \cite{cgrav-1989, cgrav-2012}. The EOMs of this theoretical framework are derived from the following gravitational action:

\begin{align} S_\text{grav}^W=\alpha\int d^4x\sqrt{-g}\,C^2,\label{cgrav-action}\end{align} where $\alpha$ is a dimensionless constant and $C^2\equiv C_{\mu\lambda\nu\sigma}C^{\mu\lambda\nu\sigma}$ ($C_{\mu\lambda\nu\sigma}$ is the Weyl conformal tensor). The EOMs read (we include the coupling of matter fields);

\begin{align} W^{(2)}_{\mu\nu}-\frac{1}{3}\,W^{(1)}_{\mu\nu}=\frac{1}{4\alpha}\,T^{(m)}_{\mu\nu},\label{cgrav-eom}\end{align} where the following tensors have been defined:

\begin{align} W^{(1)}_{\mu\nu}=&2g_{\mu\nu}\nabla^2R-2\nabla_\mu\nabla_\nu R-2RR_{\mu\nu}+\frac{1}{2}\,g_{\mu\nu}R^2,\label{w1}\\
W^{(2)}_{\mu\nu}=&\frac{1}{2}\,g_{\mu\nu}\nabla^2R+\nabla^2R_{\mu\nu}+\frac{1}{2}\,g_{\mu\nu}R_{\lambda\sigma}R^{\lambda\sigma}-2R_{\mu\lambda}R^\lambda_{\;\;\nu}-\nabla^\lambda\nabla_\mu R_{\lambda\nu}-\nabla^\lambda\nabla_\nu R_{\lambda\mu}.\label{w2}\end{align} In this case, there is no scalar field $\phi$ involved, so our demonstration in Section \ref{sect-var} that the vanishing trace of the matter SET is not required, is not applicable. Besides, note the distinguishing property of this theory: If we take the trace of \eqref{cgrav-eom}, we get $T^{(m)}=0$. Hence, only the matter with traceless SET can be coupled to conformal gravity \eqref{cgrav-action}. This time, the traceless condition is derived directly from the EOM independent of any scalar field. The demonstration of the traceless condition is straightforward: the trace $W^{(1)}=6\nabla^2R$, while $W^{(2)}=3\nabla^2R-2\nabla^\sigma\nabla^\lambda R_{\sigma\lambda}=2\nabla^2R$, where we used the Bianchi identity $\nabla^\lambda G_{\lambda\mu}=0$. The trace of the EOM \eqref{cgrav-eom} is given by 

\begin{align} T^{(m)}=4\alpha[W^{(2)}-\frac{1}{3}\,W^{(1)}]=0.\nonumber\end{align}

Nevertheless, a Weyl-invariant gravitational action leading to a theory where GR is recovered at a certain limit or in a certain gauge may be of the form:

\begin{align} S_\text{grav}=\frac{1}{2}\int d^4x\sqrt{-g}\left[\frac{1}{6}\,\phi^2R+(\der\phi)^2+2C^2\right],\label{deser-action'}\end{align} which is called massive conformal gravity \cite{faria-b, faria-c, faria-d, faria-e, faria-f}. In this case, the demonstration in Section \ref{sect-var} is applicable again, so the traceless condition does not arise.


Interestingly, the above Weyl-invariant framework may provide a suitable classical description of the gravitational interactions of matter, which meets Dicke's principle that the (classical) laws of physics must be invariant under the transformations of units; the name used by Dicke for the Weyl transformations \eqref{weyl-t}. Frequent arguments against the Weyl invariance of physical laws must be adequately identified as arguments against local scale invariance, instead. The latter must not be confused with form-invariance under Weyl transformations:

\begin{align} g_{\mu\nu}\rightarrow\Omega^2g_{\mu\nu},\;\phi\rightarrow\Omega^{-1}\phi,\;\chi_i\rightarrow\Omega^{w_{\chi_i}}\chi_i,\;m\rightarrow\Omega^{-1}m.\nonumber\end{align}


{\bf Acknowledgments} The author thanks Felipe F. Faria for pointing out several bibliographic references and for the useful exchange of ideas and critical comments. FORDECYT-PRONACES-CONACYT has supported this work under grant CF-MG-2558591.



\section*{Declaration of generative AI and AI-assisted technologies in the manuscript preparation process.}


During the preparation of this work, the author used the WRITEFULL tool on Overleaf to check spelling and grammar. After using this tool, the author reviewed and edited the content as needed and takes full responsibility for the content of the published article.






\begin{thebibliography}{99}



\bibitem{dicke-1962} R.H. Dicke, Phys. Rev. {\bf 125} (1962) 2163-2167. \href{dicke-1962}{https://doi.org/10.1103/PhysRev.125.2163}


\bibitem{deser-1970} S. Deser, Annals Phys. {\bf 59} (1970) 248-253. \href{deser-1970}{https://doi.org/10.1016/0003-4916(70)90402-1}


\bibitem{callan_prd_1970} C.G. Callan, Jr., Phys. Rev. D {\bf 2} (1970) 1541-1547. \href{callan_prd_1970}{https://doi.org/10.1103/PhysRevD.2.1541}

\bibitem{morganstern_1970} R.E. Morganstern, Phys. Rev. D {\bf 1} (1970) 2969-2971. \href{morganstern_1970}{https://doi.org/10.1103/PhysRevD.1.2969}

\bibitem{rosen-1970} G. Rosen, Phys. Rev. D {\bf 3} (1971) 615-616. \href{rosen-1970}{https://doi.org/10.1103/PhysRevD.3.615}

\bibitem{anderson-1971} J.L. Anderson, Phys. Rev. D {\bf 3} (1971) 1689-1691. \href{anderson-1971}{https://doi.org/10.1103/PhysRevD.3.1689}

\bibitem{fujii-1974} Y. Fujii, Phys. Rev. D {\bf 9} (1974) 874-876. \href{fujii-1974}{https://doi.org/10.1103/PhysRevD.9.874}

\bibitem{bekenstein_1980} J.D. Bekenstein, A. Meisels, Phys. Rev. D {\bf 22} (1980) 1313. \href{bekenstein_1980}{https://doi.org/10.1103/PhysRevD.22.1313}


\bibitem{cheng-1988} H. Cheng, Phys. Rev. Lett. {\bf 61} (1988) 2182-2184. \href{cheng-1988}{https://doi.org/10.1103/PhysRevLett.61.2182}

\bibitem{bekenstein_1993} J.D. Bekenstein, Phys. Rev. D {\bf 48} (1993) 3641-3647. \href{bekenstein_1993}{https://doi.org/10.1103/PhysRevD.22.1313}

\bibitem{magnano_1994} G. Magnano, L.M. Sokolowski, Phys. Rev. D {\bf 50} (1994) 5039-5059. \href{magnano_1994}{https://doi.org/10.1103/PhysRevD.50.5039}

\bibitem{capozziello_1997} S. Capozziello, R. de Ritis, A.A. Marino, Class. Quant. Grav. {\bf 14} (1997) 3243-3258.\\ \href{capozziello_1997}{https://iopscience.iop.org/article/10.1088/0264-9381/14/12/010}

\bibitem{faraoni_1998} V. Faraoni, Phys. Lett. A {\bf 245} (1998) 26-30. \href{faraoni_1998}{https://doi.org/10.1016/S0375-9601(98)00387-9}

\bibitem{faraoni_rev} V. Faraoni, E. Gunzig, P. Nardone, Fund. Cosmic Phys. {\bf 20} (1999) 121. \href{faraoni_rev}{https://arxiv.org/abs/gr-qc/9811047}
 

\bibitem{faraoni_ijmpd_1999} V. Faraoni and E. Gunzig, Int. J. Theor. Phys. {\bf 38} (1999) 217-225. \href{faraoni_ijmpd_1999}{https://doi.org/10.1023/A:1026645510351}

\bibitem{quiros_prd_2000} I. Quiros, Phys. Rev. D {\bf 61} (2000) 124026. \href{quiros_prd_2000}{https://doi.org/10.1103/PhysRevD.61.124026}

\bibitem{fabris_2000} J.C. Fabris and R. de Sa Ribeiro, Gen. Rel. Grav. {\bf 32} (2000) 2141-2158. \href{fabris_2000}{https://doi.org/10.1023/A:1001998419345}


\bibitem{fujii_book} Y. Fujii, K.I. Maeda, The Scalar-Tensor Theory of Gravitation, Cambridge University Press, Cambridge, UK, 2004.

\bibitem{faraoni_book} V. Faraoni, Cosmology in scalar tensor gravity, Kluwer Academic Publishers, The Netherlands, 2004.


\bibitem{flanagan_cqg_2004} E.E. Flanagan, Class. Quant. Grav. {\bf 21} (2004) 3817-3829.\\ \href{flanagan_cqg_2004}{https://iopscience.iop.org/article/10.1088/0264-9381/21/15/N02}
 
\bibitem{faraoni_prd_2007} V. Faraoni, S. Nadeau, Phys. Rev. D {\bf 75} (2007) 023501. \href{faraoni_prd_2007}{https://doi.org/10.1103/PhysRevD.75.023501}

\bibitem{catena_prd_2007} R. Catena, M. Pietroni, L. Scarabello, Phys. Rev. D {\bf 76} (2007) 084039. \href{catena_prd_2007}{https://doi.org/10.1103/PhysRevD.76.084039}

\bibitem{nicolai-2007} K.A. Meissner, H. Nicolai, Phys. Lett. B {\bf 648} (2007) 312-317. \href{nicolai-2007}{https://doi.org/10.1016/j.physletb.2007.03.023}

\bibitem{sotiriou_ijmpd_2008} T.P. Sotiriou, V. Faraoni, S. Liberati, Int. J. Mod. Phys. D {\bf 17} (2008) 399-423.\\ \href{sotiriou_ijmpd_2008}{https://doi.org/10.1142/S0218271808012097}

\bibitem{odi-1} E. Elizalde, S. Nojiri, S.D. Odintsov, D. Saez-Gomez, V. Faraoni, Phys. Rev. D {\bf 77} (2008) 106005.\\ \href{odi-1}{https://doi.org/10.1103/PhysRevD.77.106005}

\bibitem{elizalde_grg_2010} S. Carloni, E. Elizalde, S. Odintsov, Gen. Rel. Grav. {\bf 42} (2010) 1667-1705.\\ \href{elizalde_grg_2010}{https://doi.org/10.1007/s10714-010-0936-1}


\bibitem{deruelle_2011} N. Deruelle, M. Sasaki, Springer Proc. Phys. {\bf 137} (2011) 247-260. \href{deruelle_2011}{https://doi.org/10.1007/978-3-642-19760-4\_23}

\bibitem{chiba_2013} T. Chiba, M. Yamaguchi, JCAP {\bf 10} (2013) 040.\\ \href{chiba_2013}{https://iopscience.iop.org/article/10.1088/1475-7516/2013/10/040}


\bibitem{capozziello_prd_2013} A. Stabile, An. Stabile, S. Capozziello, Phys. Rev. D {\bf 88} (2013) 124011. \href{capozziello_prd_2013}{https://doi.org/10.1103/PhysRevD.88.124011}
 
\bibitem{quiros_grg_2013} I. Quiros, R. Garcia-Salcedo, J.E. Madriz Aguilar, T. Matos, Gen. Rel. Grav. {\bf 45} (2013) 489-518.\\ \href{quiros_grg_2013}{https://doi.org/10.1007/s10714-012-1484-7}


\bibitem{bars-2014} I. Bars, P. Steinhardt, N. Turok, Phys. Rev. D {\bf 89} (2014) 043515. \href{bars-2014}{https://doi.org/10.1103/PhysRevD.89.043515}


\bibitem{alvarez-2015} E. Álvarez, S. González-Martin, M. Herrero-Valea, JCAP {\bf 03} (2015) 035.\\ \href{alvarez-2015}{https://iopscience.iop.org/article/10.1088/1475-7516/2015/03/035}


\bibitem{jarv_2015} L. Järv, P. Kuusk, M. Saal, O. Vilson, Phys. Rev. D {\bf 91} (2015) 024041. \href{jarv_2015}{https://doi.org/10.1103/PhysRevD.91.024041}


\bibitem{jackiw-2015} R. Jackiw and S.Y. Pi, Phys. Rev. D {\bf 91} (2015) 067501. \href{jackiw-2015}{https://doi.org/10.1103/PhysRevD.91.067501}


\bibitem{sasaki_2016} G. Domènech, M. Sasaki, Int. J. Mod. Phys. D {\bf 25} (2016) 1645006. \href{sasaki_2016}{https://doi.org/10.1142/S0218271816450061}

\bibitem{banerjee_2016} N. Banerjee, B. Majumder, Phys. Lett. B {\bf 754} (2016) 129-134. \href{banerjee_2016}{https://doi.org/10.1016/j.physletb.2016.01.022}

\bibitem{pandey_2017} S. Pandey and N. Banerjee, Eur. Phys. J. Plus {\bf 132} (2017) 107. \href{pandey_2017}{https://doi.org/10.1140/epjp/i2017-11385-0}


\bibitem{quiros_ijmpd_2019} I. Quiros, Int. J. Mod. Phys. D {\bf 28} (2019) 1930012. \href{quiros_ijmpd_2019}{https://doi.org/10.1142/S021827181930012X}


\bibitem{hobson-prd-2020} M.P. Hobson, A.N. Lasenby, Phys. Rev. D {\bf 102} (2020) 084040. \href{hobson-prd-2020}{https://doi.org/10.1103/PhysRevD.102.084040}

\bibitem{hobson-epjc-2022} M. Hobson, A. Lasenby, Eur. Phys. J. C {\bf 82} (2022) 585. \href{hobson-epjc-2022}{https://doi.org/10.1140/epjc/s10052-022-10531-6}


\bibitem{bamber_prd_2023} J. Bamber, Phys. Rev. D {\bf 107} (2023) 024013. \href{bamber_prd_2023}{https://doi.org/10.1103/PhysRevD.107.024013}


\bibitem{mohamedi-2024} N. Mohammedi, Class. Quant. Grav. {\bf 41} (2024) 195021.\\ \href{mohamedi-2024}{https://iopscience.iop.org/article/10.1088/1361-6382/ad7186}


\bibitem{quiros_prd_2025} I. Quiros, Phys. Rev. D {\bf 111} (2025) 064024. \href{quiros_prd_2025}{https://doi.org/10.1103/PhysRevD.111.064024}

 
\bibitem{casas_1992} J.A. Casas, J. Garcia-Bellido, M. Quiros, Class. Quant. Grav. {\bf 9} (1992) 1371-1384.\\ \href{casas_1992}{https://iopscience.iop.org/article/10.1088/0264-9381/9/5/018}


\bibitem{faraoni_2009} V. Faraoni, Phys. Rev. D {\bf 80} (2009) 124040. \href{faraoni_2009}{https://doi.org/10.1103/PhysRevD.80.124040}


\bibitem{ccj_ann_phys_1970} C.G. Callan, Jr., S.R. Coleman, R. Jackiw, Annals Phys. {\bf 59} (1970) 42-73.\\ \href{ccj_ann_phys_1970}{https://doi.org/10.1016/0003-4916(70)90394-5}


\bibitem{kastrup_1966} H.A. Kastrup, Phys. Rev. {\bf 142} (1966) 1060-1071. \href{kastrup_1966}{https://doi.org/10.1103/PhysRev.142.1060}


\bibitem{jackiw_1972} R. Jackiw, Phys. Today {\bf 25} (1972) 23-27. \href{jackiw_1972}{https://doi.org/10.1063/1.3070673}


\bibitem{everett_rmp_1957} H. Everett, Rev. Mod. Phys. {\bf 29} (1957) 454-462. \href{everett_rmp_1957}{https://doi.org/10.1103/RevModPhys.29.454}

\bibitem{wheeler} J.A. Wheeler, Rev. Mod. Phys. {\bf 29} (1957) 463-465. \href{wheeler}{https://doi.org/10.1103/RevModPhys.29.463}

\bibitem{dewitt_phys_rev_1967} B.S. DeWitt, Phys. Rev. {\bf 160} (1967) 1113-1148. \href{dewitt_phys_rev_1967}{https://doi.org/10.1103/PhysRev.160.1113}

\bibitem{dewitt} B.S. DeWitt, Phys. Today {\bf 23} (1970) 30-35. \href{dewitt}{https://doi.org/10.1063/1.3022331}

\bibitem{barvinsky_cqg_1990} A.O. Barvinsky, A.Yu. Kamenshchik, Class. Quant. Grav. {\bf 7} (1990) 2285-2293.\\ \href{barvinsky_cqg_1990}{https://iopscience.iop.org/article/10.1088/0264-9381/7/12/010}

\bibitem{omnes_rmp_1992} R. Omnes, Rev. Mod. Phys. {\bf 64} (1992) 339-382. \href{omnes_rmp_1992}{https://doi.org/10.1103/RevModPhys.64.339}

\bibitem{tegmark_1998} M. Tegmark, Fortsch. Phys. {\bf 46} (1998) 855-862.\\ \href{tegmark_1998}{https://doi.org/10.1002/(SICI)1521-3978(199811)46:6/8<855::AID-PROP855>3.0.CO;2-Q}

\bibitem{garriga_prd_2001} J. Garriga, A. Vilenkin, Phys. Rev. D {\bf 64} (2001) 043511. \href{garriga_prd_2001}{https://doi.org/10.1103/PhysRevD.64.043511}

\bibitem{zurek_rmp_2003} W.H. Zurek, Rev. Mod. Phys. {\bf 75} (2003) 715-775. \href{zurek_rmp_2003}{https://doi.org/10.1103/RevModPhys.75.715}

\bibitem{tegmark_nature_2007} M. Tegmark, Nature {\bf 448} (2007) 23. \href{tegmark_nature_2007}{https://doi.org/10.1038/448023a}

\bibitem{quiros_prd_2023} I. Quiros, Phys. Rev. D {\bf 107} (2023) 104028. \href{quiros_prd_2023}{https://doi.org/10.1103/PhysRevD.107.104028}


\bibitem{faria-a} F.F. Faria, Eur. Phys. J. C {\bf 83} (2023) 81. \href{faria-a}{https://doi.org/10.1140/epjc/s10052-023-11204-8}


\bibitem{woodard-1986} N.C. Tsamis and R.P. Woodard, Annals Phys. {\bf 168} (1986) 457. \href{woodard-1986}{https://doi.org/10.1016/0003-4916(86)90040-0}

\bibitem{quiros-arxiv-2025} I. Quiros, A.K. Rao, e-Print: 2503.12826 \\ \href{quiros-arxiv-2025}{
https://doi.org/10.48550/arXiv.2503.12826}

\bibitem{brans-1988} C.H. Brans, Class. Quantum Grav. {\bf 5} (1988) L197-L199.\\ \href{brans-1988}{https://iopscience.iop.org/article/10.1088/0264-9381/5/12/001}


\bibitem{bambi-2017-a} C. Bambi, L. Modesto, L. Rachwał, JCAP {\bf 05} (2017) 003.\\ \href{bambi-2017-a}{https://iopscience.iop.org/article/10.1088/1475-7516/2017/05/003}

\bibitem{bambi-2017-b} C. Bambi, L. Modesto, S. Porey, L. Rachwał, JCAP {\bf 09} (2017) 033.\\ \href{bambi-2017-b}{https://iopscience.iop.org/article/10.1088/1475-7516/2017/09/033}

\bibitem{bambi-2017-c} C. Bambi, L. Modesto, Phys. Rev. D {\bf 95} (2017) 064006. \href{bambi-2017-c}{https://doi.org/10.1103/PhysRevD.95.064006}

\bibitem{bambi-2018-a} Q. Zhang, L. Modesto, C. Bambi, Eur. Phys. J. C {\bf 78} (2018) 506. \href{bambi-2017-c}{https://doi.org/10.1140/epjc/s10052-018-5987-6}

\bibitem{bambi-2018-b} M. Zhou, Z. Cao, A. Abdikamalov, D. Ayzenberg, C. Bambi et al., Phys. Rev. D {\bf 98} (2018) 024007.\\ \href{bambi-2018-b}{https://doi.org/10.1103/PhysRevD.98.024007}


\bibitem{ghilen-2019} D.M. Ghilencea, JHEP {\bf 03} (2019) 049. \href{ghilen-2019}{https://doi.org/10.1007/JHEP03(2019)049}

\bibitem{harko-prd-2023} P. Burikham, T. Harko, K. Pimsamarn, S. Shahidi, Phys. Rev. D {\bf 107} (2023) 064008.\\ \href{harko-prd-2023}{https://doi.org/10.1103/PhysRevD.107.064008}

\bibitem{harko-epjc-2022} J.Z. Yang, S. Shahidi, T. Harko, Eur. Phys. J. C {\bf 82} (2022) 1171. \href{harko-epjc-2022}{https://doi.org/10.1140/epjc/s10052-022-11131-0}


\bibitem{modesto-2020} Q. Li, L. Modesto, Grav. Cosmol. {\bf 26} (2020) 99-117. \href{modesto-2020}{https://doi.org/10.1134/S0202289320020085}

\bibitem{modesto-2024} L. Modesto, T. Zhou, Q. Li, Universe {\bf 10} (2024) 19. \href{modesto-2024}{https://doi.org/10.3390/universe10010019}


\bibitem{capper-duff-1974} D.M. Capper, M.J. Duff, Nuovo Cim. A {\bf 23} (1974) 173-183. \href{capper-duff-1974}{https://doi.org/10.1007/BF02748300} 

\bibitem{duff-1977} M.J. Duff, Nucl. Phys. B {\bf 125} (1977) 334-348. \href{duff-1977}{https://doi.org/10.1016/0550-3213(77)90410-2}

\bibitem{rusos-1984} E.S. Fradkin, A.A. Tseytlin, Phys. Lett. B {\bf 134} (1984) 187-193. \href{rusos-1984}{https://doi.org/10.1016/0370-2693(84)90668-3}

\bibitem{affleck-1986} I. Affleck, Phys. Rev. Lett. {\bf 56} (1986) 746-748. \href{affleck-1986}{https://doi.org/10.1103/PhysRevLett.56.746}

\bibitem{duff-1994} M.J. Duff, Class. Quant. Grav. {\bf 11} (1994) 1387-1404.\\ \href{duff-1994}{https://iopscience.iop.org/article/10.1088/0264-9381/11/6/004}

\bibitem{odintsov-2000} S. Nojiri, S.D. Odintsov, Int. J. Mod. Phys. A {\bf 15} (2000) 413-428. \href{odintsov-2000}{https://doi.org/10.1142/S0217751X00000197}

\bibitem{cai-2010} R.G. Cai, L.M. Cao, N. Ohta, JHEP {\bf 04} (2010) 082. \href{cai-2010}{https://doi.org/10.1007/JHEP04(2010)082}


\bibitem{cgrav-1989} P.D. Mannheim, D. Kazanas, Astrophys. J. {\bf 342} (1989) 635-638. \href{cgrav-1989}{https://doi.org/10.1086/167623}

\bibitem{cgrav-2012} P.D. Mannheim, Found. Phys. {\bf 42} (2012) 388-420. \href{cgrav-2012}{https://doi.org/10.1007/s10701-011-9608-6}


\bibitem{riess-1998} A.G. Riess et al (Supernova Search Team), Astron. J. {\bf 116} (1998) 1009-1038.\\ \href{riess-1998}{https://iopscience.iop.org/article/10.1086/300499}


\bibitem{perlmutter-1999} S. Perlmutter et al (Supernova Cosmology Project Collaboration), Astrophys. J. {\bf 517} (1999) 565-586.\\ \href{perlmutter-1999}{https://iopscience.iop.org/article/10.1086/307221}

  
\bibitem{desi} A.G. Adame et al. (DESI Collaboration), JCAP {\bf 02} (2025) 021.\\ \href{desi}{https://iopscience.iop.org/article/10.1088/1475-7516/2025/02/021}



\bibitem{ccp-padman} T. Padmanabhan, Phys. Rept. {\bf 380} (2003) 235-320 [e-Print: hep-th/0212290].


\bibitem{dodelson-book} S. Dodelson, ``Modern cosmology,'' (Academic Press, Amsterdam, 2003).\\  \href{dodelson-book}{https://doi.org/10.1016/B978-0-12-219141-1.X5019-0}


\bibitem{faria-b} F.F. Faria, Adv. High Energy Phys. {\bf 2014} (2014) 520259. \href{faria-b}{ https://doi.org/10.1155/2014/520259}

\bibitem{faria-c} F.F. Faria, Eur. Phys. J. C {\bf 76} (2016) 188. \href{faria-c}{https://doi.org/10.1140/epjc/s10052-016-4037-5}

\bibitem{faria-d} F.F. Faria, Eur. Phys. J. C {\bf 78} (2018) 277. \href{faria-d}{https://doi.org/10.1140/epjc/s10052-018-5762-8}

\bibitem{faria-e} F.F. Faria, Mod. Phys. Lett. A {\bf 36} (2021) 2150115. \href{faria-e}{https://doi.org/10.1142/S0217732321501157}

\bibitem{faria-f} F.F. Faria, Mod. Phys. Lett. A {\bf 37} (2022) 2250033. \href{faria-f}{https://doi.org/10.1142/S021773232250033X}


\bibitem{penrose-1} R. Penrose, Phys. Rev. Lett. {\bf 14} (1965) 57-59. \href{penrose-1}{https://doi.org/10.1103/PhysRevLett.14.57}

\bibitem{penrose-2} R. Penrose, Riv. Nuovo Cim. {\bf 1} (1969) 252-276; Gen. Rel. Grav. {\bf 34} (2002) 1141-1165.\\ \href{penrose-2}{https://doi.org/10.1023/A:1016578408204}

\bibitem{penrose-3} S.W. Hawking, R. Penrose, Proc. Roy. Soc. Lond. A {\bf 314} (1970) 529-548. \\ \href{penrose-3}{https://doi.org/10.1098/rspa.1970.0021}

\bibitem{hawking-1} S.W. Hawking, Phys. Rev. D {\bf 14} (1976) 2460-2473. \href{hawking-1}{https://doi.org/10.1103/PhysRevD.14.2460}

\bibitem{berger} B.K. Berger, Living Rev. Rel. {\bf 5} (2002) 1. \href{berger}{https://link.springer.com/article/10.12942/lrr-2002-1}

\bibitem{bosma} L. Bosma, B. Knorr, F. Saueressig, Phys. Rev. Lett. {\bf 123} (2019) 101301.\\ \href{bosma}{https://doi.org/10.1103/PhysRevLett.123.101301}


\bibitem{wald-book} R.M. Wald, ``General Relativity,'' (The University of Chicago Press, Chicago, 1984).\\ \href{wald-book}{https://doi.org/10.7208/chicago/9780226870373.001.0001}


\bibitem{hawking-2} G.W. Gibbons, S.W. Hawking, Phys. Rev. D {\bf 15} (1977) 2752-2756. \href{hawking-2}{https://doi.org/10.1103/PhysRevD.15.2752}

\bibitem{hawking-3} S.W. Hawking, Phys. Rev. D {\bf 18} (1978) 1747-1753. \href{hawking-3}{https://doi.org/10.1103/PhysRevD.18.1747}

\bibitem{hawking-4} G.W. Gibbons, S.W. Hawking, M.J. Perry, Nucl. Phys. B {\bf 138} (1978) 141-150.\\ \href{hawking-4}{https://doi.org/10.1016/0550-3213(78)90161-X}


\bibitem{wands-rev} J.E. Lidsey, D. Wands, E.J. Copeland, Phys. Rept. {\bf 337} (2000) 343-492.\\ \href{wands-rev}{https://doi.org/10.1016/S0370-1573(00)00064-8}






\end{thebibliography}
\end{document}